\documentclass{JHEP3} 

\usepackage{amsmath}
\usepackage[dvips]{graphicx}

\providecommand{\openone}{\leavevmode\hbox{\small1\kern-3.5pt\normalsize1}}

\newcommand{\as}{\alpha_s}
\newcommand{\wh}{\widehat}
\newcommand{\nn}{\nonumber}
\newcommand{\ve}{\varepsilon}
\newcommand{\IM}{\mbox{\rm Im}}
\newcommand{\RE}{\mbox{\rm Re}}

\newcommand{\eqn}[1]{(\ref{#1})}
\newcommand{\mev}{\mbox{\rm MeV}}
\newcommand{\gev}{\mbox{\rm GeV}}
\newcommand{\tvs}{\vbox{\vskip 6mm}}
\newcommand{\smvs}{\vbox{\vskip 8mm}}

\newcommand{\MSb}{{\overline{\rm MS}}}
\newcommand{\sfrac}[2]{\mbox{$\frac{#1}{#2}$}}
\newcommand\fverb{\setbox\pippobox=\hbox\bgroup\verb}
\newcommand\fverbdo{\egroup\medskip\noindent%
                        \fbox{\unhbox\pippobox}\ }
\newcommand\fverbit{\egroup\item[\fbox{\unhbox\pippobox}]}
\newbox\pippobox

\preprint{PITHA 08/11, UAB-FT-648\\
}%

\title{\boldmath{$\as$} and the \boldmath{$\tau$} hadronic width: 
fixed-order, contour- improved and higher-order perturbation theory}

\author{Martin Beneke\\
        Institut f\"ur Theoretische Physik E, RWTH Aachen University,\\ 
        D-52056 Aachen,
        Germany
       }

\author{Matthias Jamin\\
        Instituci\'o Catalana de Recerca i Estudis Avan\c{c}ats (ICREA),\\
        IFAE, Theoretical Physics Group, UAB,
        E-08193 Bellaterra, Barcelona, Spain\\
        E-mail: \email{jamin@ifae.es}}

\abstract{
The determination of $\as$ from hadronic $\tau$ decays is revisited, with a
special emphasis on the question of higher-order perturbative corrections and
different possibilities of resumming the perturbative series with the
renormalisation group: fixed-order (FOPT) vs. contour-improved perturbation
theory (CIPT). The difference between these approaches has evolved into a
systematic effect that does not go away as higher orders in the perturbative
expansion are added. We attempt to clarify under which circumstances one or
the other approach provides a better approximation to the true result. To this
end, we propose to describe the Adler function series by a model that includes
the exactly known coefficients and theoretical constraints on the large-order
behaviour originating from the operator product expansion and the
renormalisation group. Within this framework we find that while CIPT is unable
to account for the fully resummed series, FOPT smoothly approaches the Borel
sum, before the expected divergent behaviour sets in at even higher orders.
Employing FOPT up to the fifth order to determine $\as$ in the $\MSb$ scheme,
we obtain $\as(M_\tau)=0.320 {}^{+0.012}_{-0.007}$, corresponding to 
$\as(M_Z) = 0.1185 {}^{+0.0014}_{-0.0009}$.  Improving this result by including
yet higher orders from our model yields $\as(M_\tau)=0.316 \pm 0.006$, which
after evolution leads to $\as(M_Z) = 0.1180 \pm 0.0008$. Our results are lower
than previous values obtained from $\tau$ decays.
}%

\keywords{QCD, hadronic $\tau$ decays, perturbative series, renormalisation group}

\begin{document}

\section{Introduction}\label{sect1}

Precision determinations of fundamental parameters within the Standard Model
are of utmost importance in order to test its internal consistency or point
towards physics which goes beyond it. In this respect the central parameter
of the strong interaction sector is the strong coupling $\as$, and until now
tremendous efforts have been put into an ever better determination of $\as$
\cite{pdg06,bet06}.

One of the most precise determinations of $\as$, competitive with the current
world average, is provided by detailed investigations of the $\tau$ hadronic
width
\begin{equation}
\label{Rtauex}
R_\tau \,\equiv\, \frac{\Gamma[\tau^- \to {\rm hadrons} \, \nu_\tau (\gamma)]}
{\Gamma[\tau^- \to e^- \overline \nu_e \nu_\tau (\gamma)]} \,=\,
3.640 \pm 0.010 \,,
\end{equation}
as well as invariant mass distributions
\cite{bnp92,cleo95,aleph98,opal99,aleph05}. The recent analyses of the ALEPH
spectral function data \cite{aleph05,dhz05,ddhmz08} on the basis of the final
full LEP data set yielded
$\as(M_\tau)=0.344\pm 0.005_{\rm exp}\pm 0.007_{\rm th}$ which after evolution
to the $Z$-boson mass scale results in $\as(M_Z)=0.1212\pm 0.0011$. The
dominant quantifiable theory uncertainty resides in the contribution of as yet
uncalculated higher-order perturbative QCD corrections and improvements of the
perturbative series through renormalisation group methods.

Of particular interest for the $\as$ determination is the $\tau$ decay rate
into light $u$ and $d$ quarks, $R_{\tau,V/A}$, which proceeds either through a 
vector or an axialvector current, since in this case power corrections are
especially suppressed. Theoretically, $R_{\tau,V/A}$ can be expressed in the
form \cite{bnp92,bra88,np88,bra89}
\begin{equation}
\label{RtauVA}
R_{\tau,V/A} \,=\, \frac{N_c}{2}\,S_{\rm EW}\,|V_{ud}|^2\,\biggl[\,
1 + \delta^{(0)} + \delta_{\rm EW}' + \sum\limits_{D\geq 2} 
\delta_{ud,V/A}^{(D)} \,\biggr] \,.
\end{equation}
Here, $S_{\rm EW}=1.0198\pm 0.0006$ \cite{ms88} and
$\delta_{\rm EW}'=0.0010\pm 0.0010$ \cite{bl90} are electroweak corrections,
$\delta^{(0)}$ comprises the perturbative QCD correction which will be our
main interest in this work, and the $\delta_{ud,V/A}^{(D)}$ denote quark mass
and higher $D$-dimensional operator corrections which arise in the framework
of the operator product expansion (OPE). The higher-order OPE contributions
are small and will only be considered towards the end of our work, when we
present our determination of $\as$.  

A particular problem emerges from the observation that different ways of
performing the renormalisation group resummation, namely fixed-order (FOPT)
or contour-improved perturbation theory (CIPT) \cite{piv91,dp92a}, apparently
lead to differing results. This is especially noteworthy as historically the
values of $\as$ extracted from $\tau$ decays employing CIPT have always been
on the high side of the world averages, and with the recent update
$\as(M_Z)=0.1185\pm 0.0010$ \cite{bet06} of the latter, this disparity is
becoming significant.

CIPT is conventionally the method of choice, since the expansion of the running
coupling $\as(\sqrt{s})$ in $\as(M_\tau)$ used in FOPT within a certain
contour integral (see section~\ref{sect3}) is near its radius of convergence
and thus argued to lead to a poorly behaved fixed-order series. This argument,
however, is not entirely compelling, since QCD perturbation series have zero
radius of convergence anyway, and are asymptotic at best, no matter whether
CIPT or FOPT is used. Indeed, in the large-$\beta_0$ approximation, which
may be viewed as a toy model for the entire perturbation series, FOPT was
identified to provide the better approximation to the full result \cite{bbb95}.
With the recent calculation of the ${\cal O}(\as^4)$ term in the series 
expansion of the Adler function \cite{bck08}, the discrepancy between FOPT
and CIPT appears to be the largest systematic theoretical uncertainty of the
$\as$ determination, as it is evident that it does not go away by adding
the presently known higher-order terms.

The following study is motivated by the need to resolve this discrepancy and
to understand its origin. Previous investigations \cite{bbb95,jam05} show that
a preference for CIPT or FOPT may strongly depend on the assumptions made on
higher-order terms in the perturbation expansion. Thus, in section~\ref{sect4},
we study several toy models in order to address, for each model, the following
questions:
\begin{itemize}

\item[i)] Are FO and CI perturbation theory seen to be compatible, once terms
beyond the currently known coefficients of the perturbative series for
$\delta^{(0)}$ are included?

\item[ii)] How do FO and CI perturbation theory at a particular order compare
to the true result for $\delta^{(0)}$, and is the closest approach
to the true  result related to the minimal terms in the respective series?

\item[iii)] And finally, which of the two methods, FOPT or CIPT provides the
closer approach to the true value at order ${\cal O}(\as^4)$, and in
general?

\end{itemize}
Here our working assumption is that the true result is approximated with
reasonable accuracy by the Borel sum of the model series, since the power
corrections to $R_\tau$ are known to be small.

The lessons learnt from the toy models lead us to proposing an ansatz for the
Adler function series, more precisely its Borel transform, which incorporates
the presently available terms in the perturbative expansion, as well as known
features of renormalon singularities \cite{ben98} determined solely by the
operator product expansion and the renormalisation group. This ansatz is
described and analysed in detail in section~\ref{sect6}. While in the toy
models discussed in section~\ref{sect4}, one may obtain compatible descriptions
of the perturbative series by FOPT or CIPT, or a preference for one of the two
prescriptions, we find that the features favouring FOPT prevail in our ansatz
for the physical case. For the physical model, CIPT never comes close to the
result for the Borel sum. On the other hand, FOPT approaches this sum in a
smooth fashion until its minimal term, after which the expected asymptotic
(divergent) behaviour sets in.

We believe that these features are characteristic to $R_\tau$ and therefore
argue that FOPT provides the better approximation to the perturbative series 
for $\delta^{(0)}$, both at ${\cal O}(\as^4)$ and in general. Based on
this observation we proceed to determine the strong coupling $\as$ in
section~\ref{sect7} in two ways: first, employing FOPT and an estimate of the 
${\cal O}(\as^5)$ term suggested by several independent arguments; second,
employing our ansatz for the entire series as discussed in section~\ref{sect6}.
Both approaches lead to values of $\as(M_Z)$ systematically lower than
previous determinations from hadronic $\tau$ decays employing CIPT.

\section{Theoretical framework}\label{sect2}

We briefly review the main theoretical expressions required in the
analysis of the inclusive hadronic $\tau$ decay width. Further details and
complete expressions can be found in the original works \cite{bnp92,pp98,pp99}.
The central quantities in such an analysis are the two-point correlation
functions
\begin{equation}
\label{PiVAmunu}
\Pi_{\mu\nu,ij}^{V/A}(p) \,\equiv\,  i\!\int \! dx \, e^{ipx} \,
\langle\Omega|\,T\{ J_{\mu,ij}^{V/A}(x)\,J_{\nu,ij}^{V/A}(0)^\dagger\}|
\Omega\rangle\,,
\end{equation}
where $|\Omega\rangle$ denotes the physical vacuum and the hadronic
vector/axialvector currents are given by
$J_{\mu,ij}^{V/A}(x)=[\bar q_j\gamma_\mu(\gamma_5)q_i](x)$. The indices $i,j$
stand for the light quark flavours up, down and strange. The correlators
$\Pi_{\mu\nu,ij}^{V/A}(p)$ have the Lorentz decomposition
\begin{equation}
\Pi_{\mu\nu,ij}^{V/A}(p) \,=\, (p_\mu p_\nu-g_{\mu\nu}p^2)\,\Pi^{V/A,(1)}_{ij}
(p^2) + p_\mu p_\nu\,\Pi^{V/A,(0)}_{ij}(p^2) \,,
\end{equation}
where the superscripts denote the components corresponding to angular momentum
$J=1$ (transversal) and $J=0$ (longitudinal) in the hadronic rest frame.

Experimentally, the hadronic decay rate of the $\tau$ lepton can be separated
into the contributions of vector $R_{\tau,V}$ and axialvector $R_{\tau,A}$
components for the $(\bar ud)$-quark current as well as the contribution with
net-strangeness $R_{\tau,S}$, resulting from the $(\bar us)$-quark
current,
\begin{equation}
\label{RTauex}
R_\tau \,=\, R_{\tau,V} + R_{\tau,A} + R_{\tau,S} \,.
\end{equation}
In the Cabibbo-suppressed ($\bar us$) sector, a separation of vector from
axialvector contributions is problematic, since $G$-parity is not a good
quantum number in modes with strange particles.\footnote{A small component
with strange quarks but without net-strangeness also resides in $R_{\tau,V}$
and $R_{\tau,A}$, with the dominant decay channel being
$\tau^-\to\pi^-K^0\bar K^0\nu_\tau$.} On the theory side, $R_\tau$ can be
expressed as an integral of the spectral functions $\IM\,\Pi^{(1)}(s)$ and
$\IM\,\Pi^{(0)}(s)$ over the invariant mass $s=p^2$ of the final state
hadrons~\cite{tsai71}:
\begin{equation}
\label{Rtauth}
R_\tau \,=\, 12\pi \!\int\limits_0^{M_\tau^2} \frac{ds}{M_\tau^2}\,\biggl(
1-\frac{s}{M_\tau^2}\biggr)^{\!2} \biggl[\biggl(1+2\frac{s}{M_\tau^2}\biggr)
\IM\,\Pi^{(1)}(s)+\IM\,\Pi^{(0)}(s) \,\biggr] \,.
\end{equation}
For simplicity, in the following, we shall omit the EW correction factor
$S_{\rm EW}$, but it will of course be included in our final numerical analysis
for $\as$. The appropriate combinations of the two-point correlation functions
resulting from the weak decay through the $W$-boson are given by
\begin{equation}
\Pi^{(J)}(s) \,\equiv\,  |V_{ud}|^2\Big[\, \Pi^{V,(J)}_{ud}(s) +
                                           \Pi^{A,(J)}_{ud}(s) \,\Big] +
                         |V_{us}|^2\Big[\, \Pi^{V,(J)}_{us}(s) +
                                           \Pi^{A,(J)}_{us}(s) \,\Big] \,,
\end{equation}
with $V_{ij}$ being the corresponding elements of the Cabibbo-Kobayashi-Maskawa
quark-mixing matrix.

The {\em exact} (non-perturbative) correlation functions are  analytic in the
complex $s$-plane cut along the positive axis. Exploiting this property,
eq.~\eqn{Rtauth} can be expressed as a contour integral in the complex
$s$-plane running counter-clockwise around the circle $|s|=M_\tau^2$
\cite{bnp92}
\begin{equation}
\label{Rtaucon}
R_\tau \,=\, 6\pi i \!\!\!\oint\limits_{|s|=M_\tau^2} \frac{d s}{M_\tau^2}\,
\biggl(1-\frac{s}{M_\tau^2}\biggr)^{\!2} \biggl[\, \biggl(1+2\frac{s}{M_\tau^2}
\biggr)\,\Pi^{(1)}(s) + \Pi^{(0)}(s) \,\biggr] \,.
\end{equation}
The same analytic property holds to any finite order in perturbation theory
in $\as$ (although the discontinuity is arbitrarily wrong at small $s$).
Eqs.~\eqn{Rtauth} and \eqn{Rtaucon} are equivalent if the correlation
functions are substituted either by the exact values or finite order
perturbative expansions. The equivalence of eqs.~\eqn{Rtauth} and \eqn{Rtaucon}
does not hold in renormalisation group improved perturbation theory due to the
Landau pole singularity \cite{bbb95}. 

Whereas the correlators $\Pi^{(1)}(s)$ and $\Pi^{(0)}(s)$ themselves are not
physical quantities in the sense that they contain renormalisation scale and
scheme dependent subtraction constants and thus do not satisfy homogeneous
renormalisation group equations, by means of partial integration,
eq.~\eqn{Rtaucon} can be rewritten in terms of the physical correlation
functions
$D^{(1+0)}(s)$ and $D^{(0)}(s)$,
\begin{equation}
\label{D10D0}
D^{(1+0)}(s) \,\equiv\, -\,s\,\frac{d}{ds}\,\Big[\Pi^{(1+0)}(s)\Big] \,,
\qquad
D^{(0)}(s) \,\equiv\, \frac{s}{M_\tau^2}\,\frac{d}{ds}\,\Big[s\,\Pi^{(0)}(s)
\Big] \,,
\end{equation}
the first of which being the well-known Adler function \cite{adl74}. The
renormalisation dependent contributions to $\Pi^{(1)}(s)$ and $\Pi^{(0)}(s)$
drop out after contour integration, and in FOPT the perturbative expansions
for $R_\tau$ based on $\Pi(s)$ or $D(s)$ are identical. However, after RG
improvement in CIPT, for the first few terms the series displays a faster rate
of convergence when employing the correlators $D^{(1+0)}(s)$ and $D^{(0)}(s)$.
Thus in the present work we shall only consider an analysis of $R_\tau$ based
on this choice. Utilising the dimensionless integration variable
$x\equiv s/M_\tau^2$, eq.~\eqn{Rtaucon} then becomes
\begin{equation}
\label{Rtauklth}
R_\tau \,=\, -\,i\pi\!\oint\limits_{|x|=1} \frac{dx}{x}\,(1-x)^3
\Big[\, 3\,(1+x)\,D^{(1+0)}(M_{\tau}^2 x) + 4\,D^{(0)}(M_{\tau}^2 x)
\,\Big] \,.
\end{equation}

For large enough negative $s$, the contributions to $D^{(J)}(s)$ can be
organised in the framework of the operator product expansion in a series
of local gauge-invariant operators of increasing dimension times appropriate
inverse powers of $s$. This expansion is expected to be well behaved along the
complex contour $|s|=M_\tau^2$, except close to the crossing point with the
positive real axis~\cite{pqw76}. As can be seen from eq.~\eqn{Rtauklth},
however, the contribution near the physical cut at $s=M_\tau^2$ is strongly
suppressed by a zero of order three. Therefore, uncertainties associated with
the use of the OPE near the time-like axis are expected to be very small.
Inserting the OPE series for $D^{(J)}(s)$ into \eqn{Rtauklth}, performing the
contour integration, and extracting the terms proportional to $|V_{ud}|^2$, 
$R_{\tau,V/A}$ in the form of eq.~\eqn{RtauVA} emerges.

The purely perturbative correction $\delta^{(0)}$ only receives contributions
from the vector and axialvector correlation function in the chiral limit.
Since in this limit vector and axial-vector contributions coincide, and
$D^{(0)}(s)=0$, to investigate $\delta^{(0)}$ we can restrict ourselves
to the study of the perturbative expansion of the vector correlator
$\Pi^{(1+0)}_V(s)$ in the massless case. It exhibits the general structure
\begin{equation}
\label{Pis}
\Pi^{(1+0)}_V(s) \,=\, -\,\frac{N_c}{12\pi^2} \sum\limits_{n=0}^\infty a_\mu^n
\sum\limits_{k=0}^{n+1} c_{n,k}\,L^k  \,, \quad L\,\equiv\,
\ln\frac{-s}{\mu^2} \,,
\end{equation}
with $a_\mu\equiv a(\mu^2) \equiv\as(\mu)/\pi$ 
and $\mu$ the renormalisation scale.
As was already remarked above, $\Pi^{(1+0)}_V(s)$ itself is not a physical
quantity. However, the spectral function is as well as the Adler function
$D^{(1+0)}_V(s)$, whose general expansion then takes the form:
\begin{equation}
\label{Ds}
D^{(1+0)}_V(s) \,=\, \frac{N_c}{12\pi^2} \sum\limits_{n=0}^\infty a_\mu^n
\sum\limits_{k=1}^{n+1} k\, c_{n,k}\,L^{k-1} \,.
\end{equation}

In this expression, only the coefficients $c_{n,1}$ have to be considered as
independent. The coefficients $c_{n,k}$ with $k=2,\ldots,n+1$ can be related to
the $c_{n,1}$ and $\beta$-function coefficients by means of the renormalisation
group equation (RGE), while the coefficients $c_{n,0}$ do not appear in
measurable quantities and $c_{n,n+1}=0$ for $n\geq 1$. Up to order
$\as^4$, the RG constraints lead to:
\begin{eqnarray}
\label{cnk}
c_{2,2} &=& -\,\frac{\beta_1}{4}\,c_{1,1} \,, \quad
c_{3,3} \,=\, \frac{\beta_1^2}{12}\,c_{1,1} \,, \quad
c_{3,2} \,=\, -\,\frac{1}{4}\,(\beta_2\,c_{1,1}+2\beta_1\,c_{2,1}) \,,
\nn \\
\smvs
c_{4,4} &=& -\,\frac{\beta_1^3}{32}\,c_{1,1} \,,\quad
c_{4,3} \,=\,
\frac{\beta_1}{24}\,(5\beta_2\,c_{1,1}+6\beta_1\,c_{2,1})\,,\nn \\
\smvs
c_{4,2} &=&-\,\frac{1}{4}\,(\beta_3\,c_{1,1}+2\beta_2\,c_{2,1}+
3\beta_1\,c_{3,1}) \,. 
\end{eqnarray}
In our convention, the QCD $\beta$-function is defined as 
$\beta(a_\mu)\equiv -\,\mu\,da_\mu/d\mu = \sum_{k=1} \beta_k a_\mu^{k+1}$, with
the first coefficient being $\beta_1=11 N_c/6-N_f/3$. Since the Adler function
$D^{(1+0)}_V(s)$ satisfies a homogeneous RGE, the logarithms in eq.~\eqn{Ds}
can be summed with the choice $\mu^2=-s\equiv Q^2$, leading to the simple
expression:
\begin{equation}
\label{Dsresum}
D^{(1+0)}_V(Q^2) \,=\, \frac{N_c}{12\pi^2} \sum\limits_{n=0}^\infty
c_{n,1}\,a_Q^n \,,
\end{equation}
where $a_Q\equiv\as(Q)/\pi$. 

Until recently, the independent coefficients $c_{n,1}$ were known analytically
up to order $\as^3$ \cite{gkl91,ss91}. At $N_c=3$ in the $\MSb$-scheme
\cite{bbdm78} they read:
\begin{eqnarray}
\label{cn1}
c_{0,1} &=& c_{1,1} \,=\, 1 \,, \quad
c_{2,1} \,=\, \sfrac{365}{24}-11\zeta_3-
\Big(\sfrac{11}{12}-\sfrac{2}{3}\zeta_3\Big) N_f \,=\, 1.640 \,, \\
\smvs
c_{3,1} &=& \sfrac{87029}{288}-\sfrac{1103}{4}\zeta_3+\sfrac{275}{6}
\zeta_5 - \Big(\sfrac{7847}{216}-\sfrac{262}{9}\zeta_3+\sfrac{25}{9}\zeta_5
\Big) N_f + \Big(\sfrac{151}{162}-\sfrac{19}{27}\zeta_3\Big) N_f^2
\,=\, 6.371 \,, \nn
\end{eqnarray}
where numerical values are given at $N_f=3$. For the next five- and six-loop
coefficients $c_{4,1}$ and $c_{5,1}$, estimates employing principles of
``minimal sensitivity'' (PMS) or ``fastest apparent convergence'' (FAC)
\cite{ste81,pen82}, together with known terms of order $\as^4\,N_f^2$,
exist, which for $N_f=3$ yield \cite{ks95,bck02}:
\begin{equation}
\label{c41c51}
c_{4,1} \,=\,  27 \pm  16 \,, \quad c_{5,1} \,=\, 145 \pm 100 \,.
\end{equation}
However, as of this year, the complete result for the ${\cal O}(\as^4)$
coefficient $c_{4,1}$ is available \cite{bck08}, which greatly helps in our
analysis. At $N_f=3$ it reads:
\begin{equation}
\label{c41}
c_{4,1} \,=\, \sfrac{78631453}{20736} - \sfrac{1704247}{432}\zeta_3 +
\sfrac{4185}{8}\zeta_3^2 + \sfrac{34165}{96}\zeta_5 - \sfrac{1995}{16}\zeta_7
\,=\, 49.076 \,.
\end{equation}
Since this result turns out to be larger than the estimate presented in
eq.~\eqn{c41c51}, we shall not use the PMS/FAC prediction for $c_{5,1}$ of
\eqn{c41c51}. Instead, we attempt to estimate this coefficient based either
on a uniform convergence rate of the series, or on our model.

\section{Renormalisation group summation}\label{sect3}

We now discuss the renormalisation group improvement of the purely perturbative
correction $\delta^{(0)}$ to $R_\tau$ by means of resummation of the logarithms
appearing in eq.~\eqn{Ds}. Returning to eq.~\eqn{Rtauklth} and inserting the
general expansion \eqn{Ds} for $D^{(1+0)}_V(s)$, $\delta^{(0)}$ is found to
take the form
\begin{equation}
\label{del0}
\delta^{(0)} \,=\, \sum\limits_{n=1}^\infty a_\mu^n \sum\limits_{k=1}^{n}
k\,c_{n,k} \;\frac{1}{2\pi i}\!\!\oint\limits_{|x|=1} \!\! \frac{dx}{x}\,
(1-x)^3\,(1+x) \ln^{k-1}\biggl(\frac{-M_\tau^2 x}{\mu^2}\biggr) \,,
\end{equation}
where the identical contribution from the axialvector correlator has already
been taken into account.

As discussed above, the Adler function $D^{(1+0)}_V(s)$ and therefore also
$\delta^{(0)}$ satisfy a homogeneous RGE. In fixed-order perturbation theory
(FOPT) the logarithms in eq.~\eqn{del0} are summed by setting $\mu^2=M_\tau^2$,
leading to
\begin{equation}
\label{del0FO}
\delta^{(0)}_{\rm FO} \,=\, \sum\limits_{n=1}^\infty a(M_\tau^2)^n
\sum\limits_{k=1}^{n} k\,c_{n,k}\,J_{k-1} \,.
\end{equation}
The contour integrals $J_l$ in eq.~\eqn{del0FO} are defined by
\begin{equation}
\label{Jl}
J_l \,\equiv\, \frac{1}{2\pi i} \!\!\oint\limits_{|x|=1} \!\!
\frac{dx}{x}\, (1-x)^3\,(1+x) \ln^l(-x) \,=\,
\frac{1}{2\pi}\,\Big[\, I_{l,0} + 2\,I_{l,1} - 2\,I_{l,3} - I_{l,4} \,\Big] \,.
\end{equation}
The integrals $I_{l,m}$ are given by
\begin{eqnarray}
\label{Ilm}
I_{l,m} &\equiv&
i\!\!\!\oint\limits_{|x|=1} \!\!dx\,(-x)^{m-1} \ln^l(-x) =
i^l \!\!\int\limits_{-\pi}^{+\pi} d\alpha\,\alpha^l\,{\rm e}^{im\alpha} \,\,=
i\,\biggl(\frac{-1}{m}\biggr)^{\!l+1}\Gamma(l+1,-i\alpha m)\biggr|_{-\pi}
^{+\pi} \nn \\
\smvs
&=& (-1)^{l+m}\frac{2\,l!}{m^{l+2}}\!\sum\limits_{k=1}^{[(l+1)/2]}
(-1)^k \,\frac{m^{2k} \pi^{2k-1}}{(2k-1)!} \,,
\end{eqnarray}
where $\Gamma(l+1,z)$ is the incomplete $\Gamma$-function, $[n]$ denotes the
integer part of $n$ and $m\geq1$. For $m=0$, one obtains
$I_{l,0}\,=\,i^l[1+(-1)^l]\,\pi^{l+1}/(l+1)$. The first few of the integrals
$J_l$, which are needed up to order $\as^4$, read:
\begin{equation}
\label{J0to3}
J_0 \,=\, 1 \,, \quad
J_1 \,=\, -\,\sfrac{19}{12} \,, \quad
J_2 \,=\, \sfrac{265}{72} - \sfrac{1}{3}\,\pi^2 \,, \quad
J_3 \,=\, -\,\sfrac{3355}{288} + \sfrac{19}{12}\,\pi^2 \,,
\end{equation}
in agreement with ref.~\cite{dp92a}.

At order $\as^n$ FOPT contains unsummed logarithms of order 
$\ln^l(-x)\sim \pi^l$ with $l<n$ related to the contour integrals $J_l$.
Contour-improved perturbation theory (CIPT) sums these logarithms with the
choice $\mu^2=-M_\tau^2 x$ in eq.~\eqn{del0}, which yields
\begin{equation}
\label{del0CI}
\delta^{(0)}_{\rm CI} \,=\, \sum\limits_{n=1}^\infty c_{n,1}\,
J_n^a(M_\tau^2)
\end{equation}
in terms of the contour integrals $J_n^a(M_\tau^2)$ over the running coupling,
defined as:
\begin{equation}
\label{Jna}
J_n^a(M_\tau^2) \,\equiv\, \frac{1}{2\pi i} \!\!\oint\limits_{|x|=1}\!\!
\frac{dx}{x}\,(1-x)^3\,(1+x)\,a^n(-M_\tau^2 x) \,.
\end{equation}
In contrast to FOPT, for CIPT each order $n$ just depends on the corresponding
coefficient $c_{n,1}$. Thus, all contributions proportional to the coefficient
$c_{n,1}$ which in FOPT appear at all perturbative orders equal or greater than
$n$ are resummed into a single term. This is related to the fact that CIPT
resums the running of the QCD coupling along the integration contour in the
complex $s$-plane as can be derived directly from eq.~\eqn{Dsresum}.

In view of our numerical analysis below, a few additional remarks are in order.
Although from the form of eq.~\eqn{del0CI}, the contour integrals
$J_n^a(M_\tau^2)$ could be considered as effective couplings, they have a
rather non-trivial dependence on the perturbative order $n$. This can be seen
from figure~\ref{fig1}, in which we display $J_n^a(M_\tau^2)/a(M_\tau^2)^n$ for
an initial value $\as(M_\tau)=0.34$ \cite{aleph05} as a function of $n$ up to
$n=30$. Up to the 7th order, $J_n^a(M_\tau^2)$ is positive and then above the
7th order turns negative. This implies that $J_7^a(M_\tau^2)$ is rather small
which is later reflected in the fact that with four-loop running of $\as$
always the 7th term in the CIPT series is found smallest. This observation
already casts some doubts on the approach of treating the CIPT series in the
sense of an asymptotic series for which quite often the optimal truncation is
provided by breaking the series at the smallest term \cite{din73}. 

\FIGURE[ht]{\includegraphics[angle=0, width=14cm]{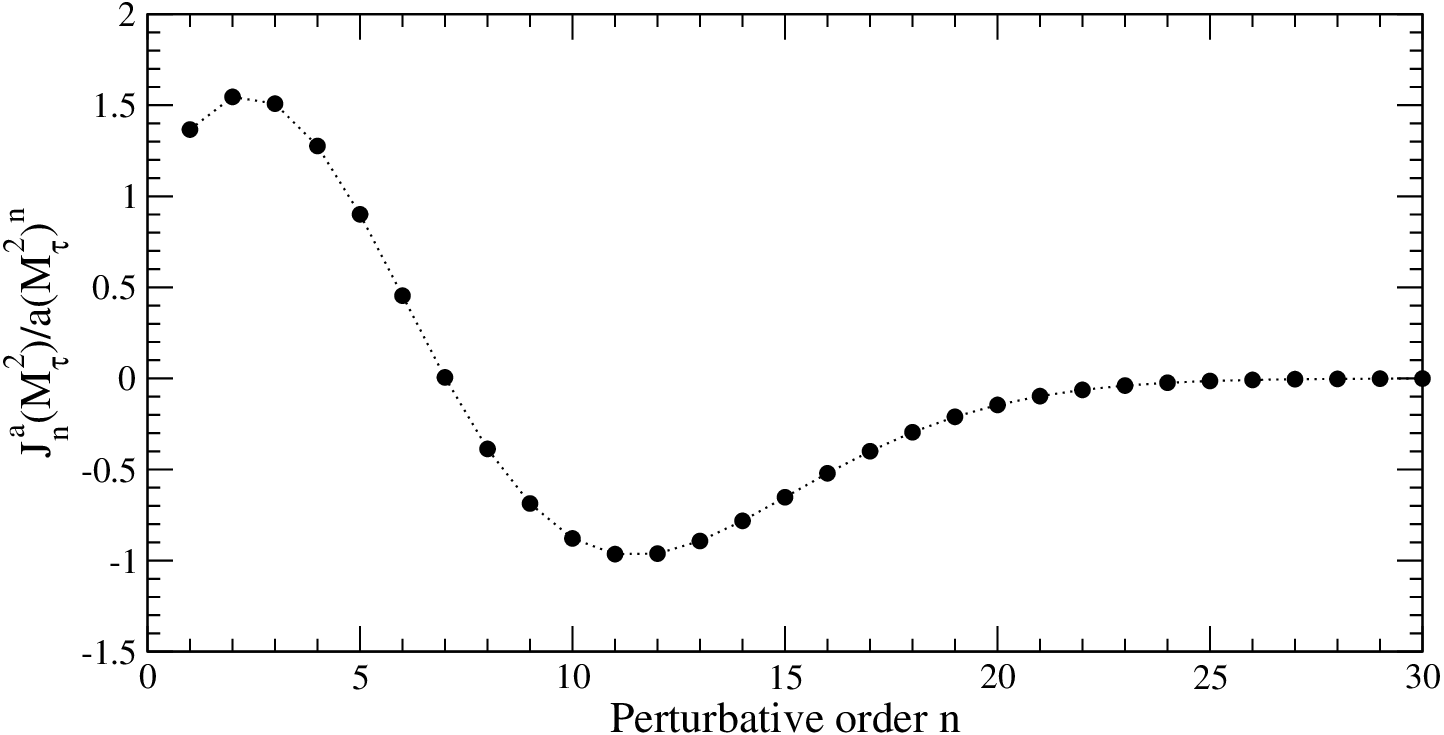}
\caption{Contour integrals $J_n^a(M_\tau^2)$ of eq.~\eqn{Jna} required for
CIPT as a function of the perturbative order $n$, computed with an initial
value $\as(M_\tau)=0.34$ and 4-loop running.\label{fig1}}}

We now recall (as is well-known, see for instance \cite{bck08}) that the two
approaches lead to significant numerical differences. Using the analytically
known coefficients of eqs.~\eqn{cn1} as well as \eqn{c41} and
$\as(M_\tau)=0.34$ 
in eqs.~\eqn{del0FO} and \eqn{del0CI}, we obtain:
\begin{eqnarray}
&& \hspace{5mm} \as^1 \hspace{11.2mm} \as^2 \hspace{11.2mm} \as^3
\hspace{11.2mm} \as^4 \hspace{12.0mm} \as^5 \nn \\
\tvs
\label{del0FOn}
\delta^{(0)}_{\rm FO} &=&
0.1082 + 0.0609 + 0.0334 + 0.0174\,(+\,0.0088\,) \,=\, 0.2200 \;(0.2288) \,, \\
\tvs
\label{del0CIn}
\delta^{(0)}_{\rm CI} &=&
0.1479 + 0.0297 + 0.0122 + 0.0086\,(+\,0.0038\,) \,=\, 0.1984 \;(0.2021) \,.
\end{eqnarray}
The CI series displays a faster convergence, but the two series do not appear
to approach a common value as successive terms are added. Summing both series
up to order $\as^4$, the difference between FO and CI perturbation theory
amounts to $0.0216$. The size of this difference is of the order of the last
included term in the FO series and about a factor of $2.5$ times the
corresponding CIPT term. This apparent disparity in the perturbative prediction
at the moment represents the dominant theoretical uncertainty in the extraction
of the strong coupling $\as$ from the hadronic $\tau$ decay rate. Thus,
it is legitimate to ask how the series for $\delta^{(0)}_{\rm CI}$ and
$\delta^{(0)}_{\rm FO}$ behave if even higher-order perturbative coefficients
are included. On the one hand both results would be expected to be compatible,
if an all-order result were available. On the other hand, re-expansion of the
contour-improved integrals $J_n^a(M_\tau^2)$ into the fixed-order series in
$a(M_\tau^2)$ results in a series that is barely convergent for realistic
values of  $a(M_\tau^2)$ \cite{dp92a}, so one might question the validity of
FOPT altogether. In the following sections we examine these issues assuming
different behaviours of the higher-order terms in the series.

An immediate question which arises is, if, on the basis of the series up to the
fourth order, we are in a position to say something about the next coefficient
$c_{5,1}$? As a somewhat naive guess, we might assume that the size of the
fifth order should be at most of the size of the previous term and larger
than zero, which is to say that the asymptotic (divergent, sign-alternating)
behaviour has not yet set in at the fifth order. Applying this criterion to
the CI series, one arrives at the estimate $0 < c_{5,1} < 642$. A slightly more
elaborate estimate can be based on the striking feature that the convergence
rate of the FOPT series is found to be very uniform. Each new term is rather
closely half of the preceeding one, and the slight dependence on the order can
even be nicely fit to a linear behaviour. Assuming that this property persists
also at the fifth order, we arrive at the estimate\footnote{Postdicting
$c_{4,1}$ in this way, we find $c_{4,1}=52\,$! The first few terms of the FO
series for $\delta^{(0)}$ are very nearly geometric.}
\begin{equation}
\label{c51}
c_{5,1} \,\approx\, 283 \,.
\end{equation}
This value lies close to the centre of the range given above, and in
section~\ref{sect6} it will be seen that eq.~\eqn{c51} is corroborated by our
model of higher order coefficients. Interestingly, it is also close to the
update of the FAC estimate to account for the newly available exact $c_{4,1}$,
which yields $c_{5,1} =275$ \cite{bck08}. Including the estimate \eqn{c51} in
the series for $\delta^{(0)}_{\rm FO}$ and $\delta^{(0)}_{\rm CI}$, the numbers
in brackets given in eqs.~\eqn{del0FOn} and \eqn{del0CIn} are obtained. Now,
the difference $\delta^{(0)}_{\rm FO} - \delta^{(0)}_{\rm CI}=0.0267$ is
increased even further, and found much larger than the last included summands.

\section{Higher orders: toy models}\label{sect4}

To acquire some feeling of what can happen to $\delta^{(0)}$ in FO and CI
perturbation theory when higher terms in the perturbative series are included,
in this section we exhibit a few toy models. Let us emphasise that we do not
believe that these models have much in common with the true QCD case (with the
exception, perhaps, of the large-$\beta_0$ approximation). Rather our concern
is to find out which features of the higher-order series determine whether
CIPT of FOPT represents a better approximation to the true result. Inspired by 
what we learn from the models, we are led to the construction of a realistic
ansatz which will be discussed in detail in the following sections.

\subsection{Truncated Adler function}

\FIGURE[ht]{\includegraphics[angle=0, width=14cm]{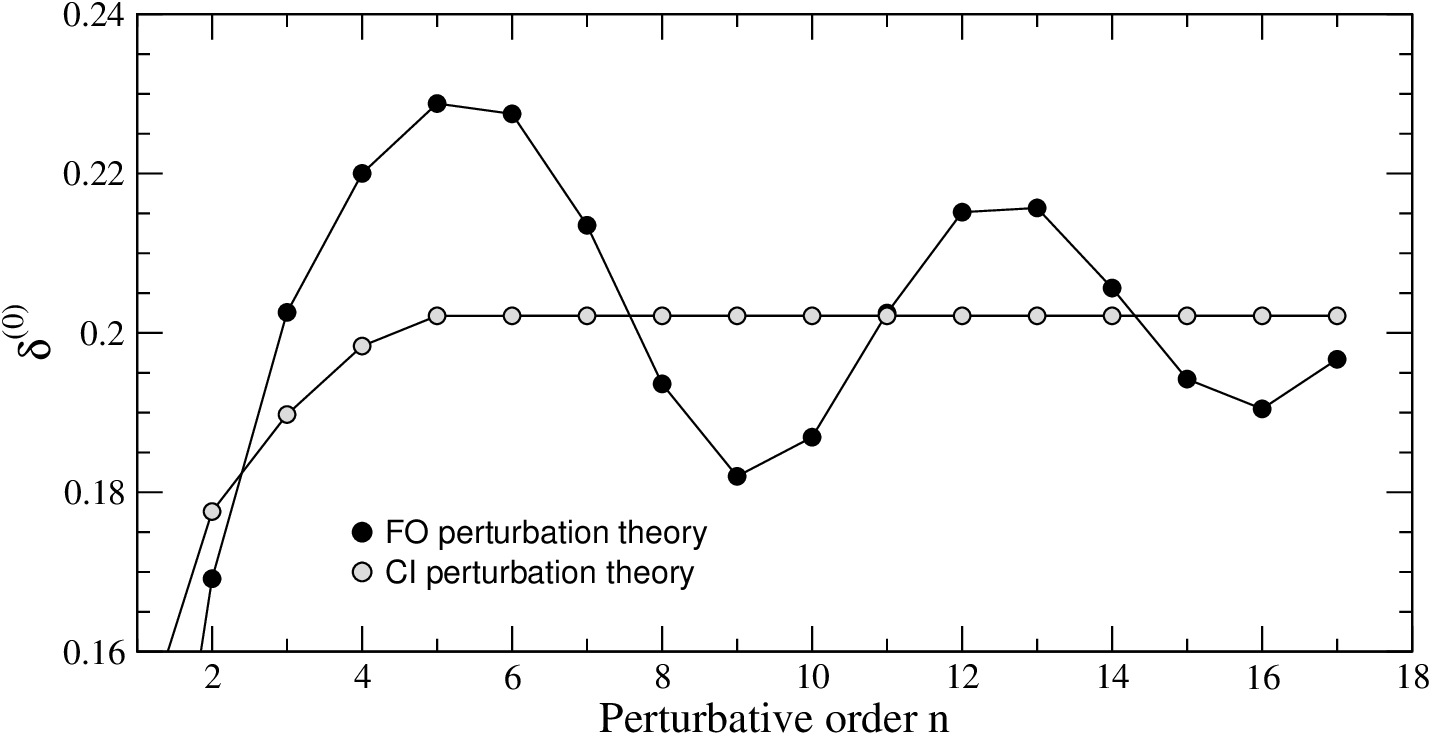}
\caption{Results for $\delta^{(0)}_{\rm FO}$ (full circles) and
$\delta^{(0)}_{\rm CI}$ (grey circles) at $\as(M_\tau)=0.34$, setting
the higher-order coefficients $c_{n,1}=0$ for $n\geq 6$. The results up to the
fifth order coincide with eqs.~\eqn{del0FOn} and \eqn{del0CIn} respectively.
\label{fig2}}}

Let us begin by considering the case in which all perturbative coefficients
$c_{n,1}$ are set to zero for $n\geq 6$. This model has already been
investigated in ref.~\cite{jam05}, in order to see how FOPT and CIPT might be
compatible even though at the fifth order their difference appears rather
dramatic.  A graphical display of the model is shown in figure~\ref{fig2}.
The result for $\delta^{(0)}_{\rm FO}$ is given as the full circles and
$\delta^{(0)}_{\rm CI}$ as the grey circles, as a function of the order up
to which the perturbative series has been summed. To guide the eye, we have
also connected the points by straight line segments. Since at each order
$\delta^{(0)}_{\rm CI}$ only depends on the explicit coefficient $c_{n,1}$,
starting from the order where the coefficients are taken to be zero, CIPT
becomes exact (and thus $\delta^{(0)}_{\rm CI}$ constant). On the contrary,
in FOPT, besides the contribution from the $c_{n,1}$, at each order we also
have contributions involving all lower $c_{k,1}$ with $k<n$, which are due to
the running of $\as$ along the complex contour. In FOPT these latter terms are
present even if the higher $c_{n,1}$ are set to zero, and entail that for
higher orders $\delta^{(0)}_{\rm FO}$ oscillates around the constant
$\delta^{(0)}_{\rm CI}$. It is evident that for the present example FOPT
represents a rather poor approximation to the exact result up to very high $n$.

Quite generally, we can write $\delta^{(0)}_{\rm FO}$ in the form  
\begin{equation}
\label{delFOcg}
\delta^{(0)}_{\rm FO} \,=\, \sum\limits_{n=1}^\infty 
\left[c_{n,1}+g_n\right] a(M_\tau^2)^n\,,
\end{equation}
where the $c_{n,1}$ series is simply the Adler function series, while the
$g_n$ series represents the additional contribution from the contour-integral
of the Adler function series. By comparison with eq.~\eqn{del0FO},
$g_n = \sum_{k=2}^n k \,c_{n,k} J_{k-1}$, hence depending only on $c_{k,1}$
with $k<n$ and $\beta$-function coefficients as remarked above. With only a
few values of $c_{n,1}$ given, the $g_n$ series has a finite radius of
convergence~\cite{dp92a}. For $\as(M_\tau)$ slightly larger than $0.34$
the series becomes divergent due to large running coupling effects along the
circular contour -- the amplitude of oscillations of FOPT around the exact
result grows and FOPT is never a good approximation. This is the reason why
CIPT is usually argued to provide the more reliable approximation to $R_\tau$. 

The general picture observed in this special case also translates to other
models in which the coefficients $c_{n,1}$ are small compared to the running
effects along the complex contour. However, the real series expansion of
$R_\tau$ is not of this general form. Rather, both the $c_n$ and $g_n$ series
are divergent with zero radius of convergence, and can at best assumed to be
asymptotic. Moreover, systematic cancellations are predicted between the $c_n$
and $g_n$ terms, when $n$ is sufficiently large.

\begin{boldmath}
\subsection{The ``large-$\beta_0$'' approximation}
\end{boldmath}

We take a first look at the issues that arise when the series have zero radius
of convergence in another toy-model, the so-called ``large-$\beta_0$''
approximation.\footnote{For historical reasons, we shall speak about the
``large-$\beta_0$'' approximation, although in the notation employed in this
work, the leading coefficient of the $\beta$-function is termed $\beta_1$. The 
``large-$\beta_0$'' approximation uses only the term with the highest power in
the number of light flavours, $N_f$, in $c_{n,1}$, and replaces $N_f$ by
$-3\beta_1$. Correspondingly, in the evolution of $\as$, only one-loop
running is taken into account.} An analytic result for the Borel transform of
the Adler function and the corresponding $R_\tau$ is available
\cite{bbb95,ben98,ben93,bro93,neu95} in this approximation, and thus the
perturbative coefficients $c_{n,1}$ are known to all orders. 

Let us briefly review the results for $\delta^{(0)}$ in the large-$\beta_0$
approximation. To make contact to the notation employed in the original works
on renormalons in connection to the Adler function and $R_\tau$
\cite{bbb95,neu95} (for a review see \cite{ben98}), it is convenient to define
the new function $\wh D(s)$ by
\begin{equation}
\label{Rs}
\frac{12\pi^2}{N_c}\,D^{(1+0)}_V(s) \,\equiv\, 1 + \wh D(s) \,\equiv\, 1 +
\sum\limits_{n=0}^\infty r_n \,\as(\sqrt{s})^{n+1} \,.
\end{equation}
Then the expansion coefficients of $D^{(1+0)}_V(s)$ and $\wh D(s)$ are related
by $c_{n,1} = \pi^n r_{n-1}$.  Next, the Borel-transform of $\wh D(s)$ is
defined by
\begin{equation}
\label{BRt}
B[\wh D](t) \,\equiv\, \sum\limits_{n=0}^\infty r_n\,\frac{t^n}{n!} \,.
\end{equation}
If $B[\wh D](t)$ has no singularities for real positive $t$ (which is not the
case for the Adler function) and does not increase too rapidly at positive
infinity, one can define the Borel integral ($\alpha$ positive) as
\begin{equation}
\label{Ralpha}
\wh D(\alpha) \,\equiv\, \int\limits_0^\infty dt\,{\rm e}^{-t/\alpha}\,
B[\wh D](t)\,,
\end{equation}
which has the same series expansion in $\alpha$ as $\wh D(s)$ does in
$\as(\sqrt{s})$.  
The integral $\wh D(\alpha)$, if it exists, gives the Borel sum
of the original divergent series. Calculating so-called bubble-chain diagrams,
it was found that the Borel-transformed Adler function $B[\wh D](t)$ obtains
infrared (IR) and ultraviolet (UV) renormalon poles at positive and negative
integer values of the variable $u\equiv\beta_1 t/(2\pi)$, respectively
\cite{ben93,bro93}. (With the exception of $u=1$.) While the IR renormalons
are related to power corrections in the operator product expansion, the
leading UV renormalon, being closest to $u=0$, dictates the large-order
behaviour of the perturbative expansion.

\TABLE[htb]{%
\begin{tabular}{ccccccc}
\hline\hline
$c_{1,1}$ & $c_{2,1}$ & $c_{3,1}$ & $c_{4,1}$ & $c_{5,1}$ & $c_{6,1}$ \\
$1$ & $1.5565$ & $15.711$ & $24.832$ & $787.83$ & $-\,1991.4$ \\
\hline
$c_{7,1}$ & $c_{8,1}$ & $c_{9,1}$ & $c_{10,1}$ & $c_{11,1}$ & $c_{12,1}$ \\
$9.857\cdot 10^4$ & $-\,1.078\cdot 10^6$ & $2.775\cdot 10^7$ &
$-\,5.388\cdot 10^8$ & $1.396\cdot 10^{10}$ & $-\,3.598\cdot 10^{11}$ \\
\hline\hline
\end{tabular}
\caption{Perturbative coefficients $c_{n,1}$ in the large-$\beta_0$
approximation up to 12th order.\label{tab1}}}

The central result of refs.~\cite{ben93,bro93} is that the Borel transform 
of the Adler function in the large-$\beta_0$ approximation (see also eq.~(5.10)
of ref.~\cite{ben98}) can be expressed as~\cite{bro93}
\begin{equation}
\label{BRuk}
B[\wh D](u) \,=\, \frac{32}{3\pi}\,
\frac{{\rm e}^{-C u}}{(2-u)}\,
\sum\limits_{k=2}^\infty\,\frac{(-1)^k k}{[k^2-(1-u)^2]^2} \,,
\end{equation}
where $C$ is a scheme-dependent constant which cancels the scheme dependence
of $\alpha$ in eq.~\eqn{Ralpha}, such that $\wh D(s)$ is independent of this
choice. In the $\MSb$-scheme it takes the value $C=-5/3$. Taylor expanding
the Borel transform in the variable $u$ and performing the Borel integral
\eqn{Ralpha} term by term, the perturbative coefficients $c_{n,1}$ for the
large-$\beta_0$ approximation can be deduced.\footnote{A particularly effective
way of analytically generating the large-$\beta_0$ coefficients $c_{n,1}$ can
be derived from eq.~(26) of ref.~\cite{bkm00}.} Numerical values for the first
12 coefficients in the $\MSb$-scheme are presented in table~\ref{tab1}. One
observes that the dominance of the leading UV renormalon at $u=-1$, and the
corresponding sign alternating behaviour of the series, in the large-$\beta_0$
approximation sets in at the 6th order.

The perturbative coefficients of table~\ref{tab1} can now be employed to
calculate $\delta^{(0)}$ in the large-$\beta_0$ approximation. For consistency
the running of $\as$ was also implemented at the one-loop order. (We shall
soon see that this plays an important role.) A graphical account of our results
at the physical coupling $\as(M_\tau)=0.34$ is displayed in
figure~\ref{fig3}. Again, the full circles correspond to
$\delta^{(0)}_{\rm FO}$ in FOPT while the grey circles provide
$\delta^{(0)}_{\rm CI}$ in the large-$\beta_0$ approximation. In both cases,
the grey diamonds represent the order at which the FO and CI series have their
smallest terms, before the asymptotic behaviour sets in. The qualitative 
picture is rather different from the previous case of the truncated Adler
function displayed in figure~\ref{fig2}. Both series appear to reach a plateau
before the divergence sets in, though the plateau is reached at higher order
for FOPT in agreement with the earlier analysis~\cite{bbb95}. However, the
difference in the plateau values of $\delta^{(0)}$ is far larger than the
minimal terms of both series, hence FOPT and CIPT seem to give incompatible 
results within the conventional uncertainty estimates.

\FIGURE[ht]{\includegraphics[angle=0, width=14cm]{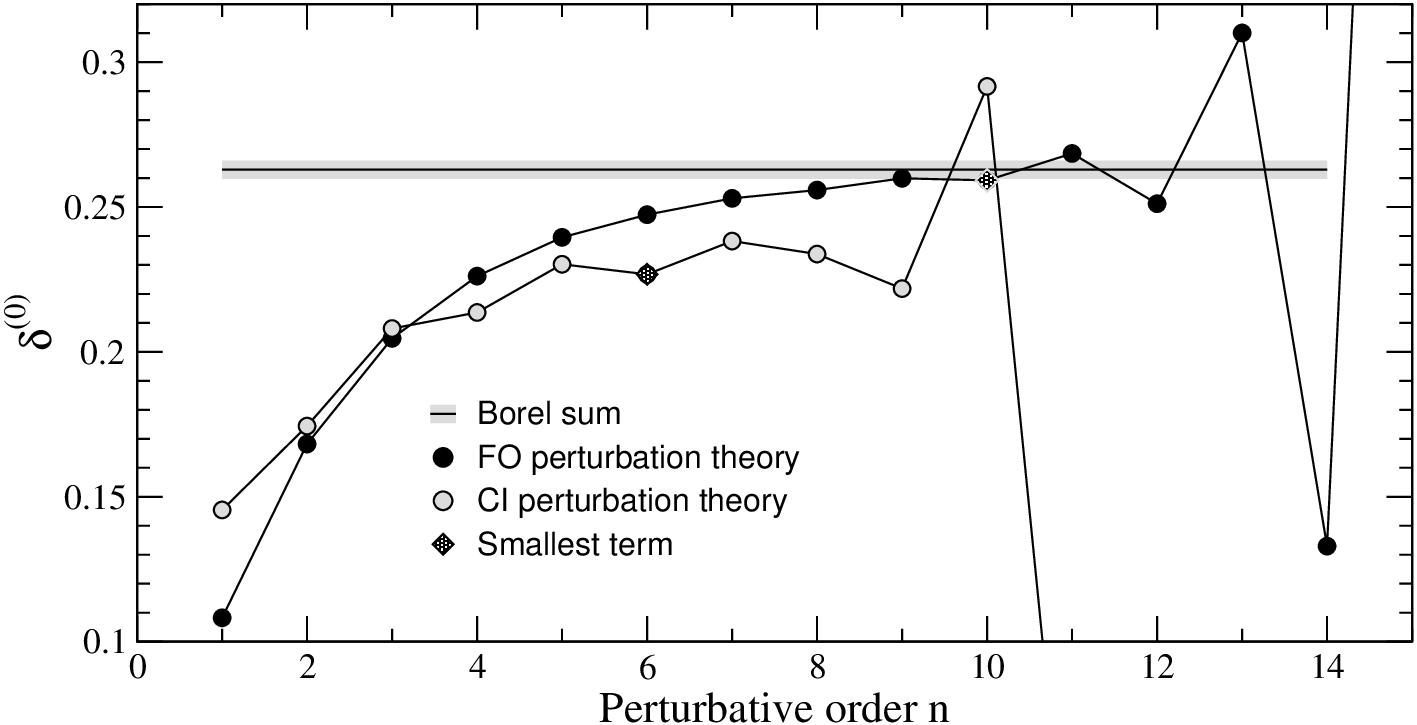}
\caption{Results for $\delta^{(0)}_{\rm FO}$ (full circles) and
$\delta^{(0)}_{\rm CI}$ (grey circles) at $\as(M_\tau)=0.34$, employing
the higher-order coefficients $c_{n,1}$ of table~\ref{tab1} obtained in the
large-$\beta_0$ approximation, as a function of the order $n$ up to which
the terms in the perturbative series have been summed. The straight line
represents the result for the Borel sum of the series, and the shaded band
provides an error estimate inferred from the complex ambiguity.\label{fig3}}}

Which is the better approximation to the ``true'' result? Since we now have
a closed expression for the Borel transform of the Adler function, we can
compute the Borel integral and compare the perturbative series to this result.
We do not expect the Borel integral to correspond exactly to the ``true''
result. First, the ``true'' result receives condensate corrections from
higher-dimensional operators in the OPE expansion. Second, when the Borel
transform has poles on the positive $t$-axis, the Borel integral must be
defined by an arbitrary deformation of the contour into the complex plane,
which introduces an ambiguity, whose dependence on $\alpha$ matches the
condensate corrections. The two are closely related (see the review
\cite{ben98}), and it has been observed~\cite{bbb95} that the size of the
ambiguity divided by $\pi$ is indeed of the order of non-perturbative 
corrections in the OPE. We thus conclude that we expect the ``true'' result to
coincide with the principal value of the Borel integral within an accuracy set
by about the ambiguity of Borel integral (divided by $\pi$), and certainly not
parametrically larger.

In figure~\ref{fig3} the horizontal line represents the value of the Borel
integral when the series is summed via the principal-value prescription, and
the shaded region marks the size of the complex ambiguity in the Borel
integral.\footnote{To obtain the maximal complex ambiguity the IR renormalon
poles are circled in such a way that they all contribute with the same sign.
The modulus of the imaginary part thus obtained is then divided by $\pi$.}
Details on the analytical calculation of the Borel transform have been
relegated to appendix~A. A good approximation of the ``true'' result should
approach the shaded region smoothly and diverge eventually. Figure~\ref{fig3}
shows the remarkable result, already observed in \cite{bbb95}, that in the
large-$\beta_0$ approximation FOPT approaches the Borel sum in a rather
monotonous fashion until the 10th order after which the sign-alternating
divergent behaviour of the series sets in, while CIPT never comes close to
the ``true'' result due to an earlier onset of the divergence. Clearly FOPT
represents the better approximation here, even at $n=4$, in stark contrast to
the previous example of the truncated Adler function series.

Let us try to gain a better insight as to why the behaviour in the
large-$\beta_0$ approximation is so different from our first model. An
immediate observation that can be made is that the $c_{n,1}$ have to be of a
similar size as the running effects $g_n$ from the contour, such that strong
cancellations between these two effects take place. To make this more precise,
consider the simple relation between $B[\wh D]$ and the correspondingly
defined Borel transform of $\delta^{(0)}$ expanded in $\as(M_\tau)$
given by~\cite{benPhD}
\begin{equation}
\label{Brtau}
B[\delta^{(0)}](u) \,=\, B[\wh D](u)\,\sin(\pi u)\left[\,\frac{1}{\pi u}+
\frac{2}{\pi (1-u)}-\frac{2}{\pi (3-u)}+\frac{1}{\pi (4-u)}\,\right] ,
\end{equation}
which is valid in the large-$\beta_0$ limit.\footnote{To derive this result,
insert eq.~\eqn{Ralpha} into eq.~\eqn{Rtauklth} and perform the contour
integral employing the one-loop expression for $\as(\sqrt{s})$.} 
We are then in a position to derive the large-order behaviour. Applying the
decomposition \eqn{delFOcg}, we obtain 
\begin{eqnarray}
\label{cgasymp}
c_{n+1,1} &=& \biggl(\frac{\beta_1}{2}\biggr)^{\!n} n!\;\biggl[\,
\phantom{-\,}\frac{4}{9} \,{\rm e}^{-5/3}\,(-1)^n \biggl(n+\frac{7}{2}\biggr) +
\frac{{\rm e}^{10/3}}{2^n} + \ldots \,\biggr] \,, \nn \\
\smvs
g_{n+1} &=& \biggl(\frac{\beta_1}{2}\biggr)^{\!n} n!\;\biggl[\,
-\,\frac{4}{9} \,{\rm e}^{-5/3}\, (-1)^n \biggl(n+\frac{16}{5}\biggr) -
\frac{{\rm e}^{10/3}}{2^n} + \ldots \,\biggr] \,,
\end{eqnarray}
where the dots denote the less important poles beyond the leading ultraviolet
renormalon pole ($u=-1$) and the leading infrared one ($u=2$). This shows
explicitly the strong cancellations that take place between the Adler function
series and the extra terms generated by the integration along the circle. The
suppression of the large-$n$ divergence allows FOPT to approach the Borel
sum smoothly. We also see why CIPT fails in this case: when such strong
cancellations between the $c_{n,1}$ and $g_n$ series are present, it is
mandatory to combine the two series at the same order. However, CIPT uses the
$c_{n,1}$ up to some finite order, while summing the $g_n$ to all orders, thus
missing the cancellation~\cite{bbb95,ben98}. The result is that CIPT runs
earlier into the leading UV renormalon divergence, as seen in figure~\ref{fig3},
though the divergence is damped by the suppression of the effective couplings
$J_n^a(M_\tau^2)$ as discussed in section~\ref{sect3}. 

There are some lessons that can be drawn from our observations which are valid
beyond the large-$\beta_0$ toy model. First, since the known exact coefficients
of the Adler function series show no interference of a sign-alternating
component, we expect the leading IR renormalon at $u=2$ to be the most relevant
contribution, before the eventual sign-alternation takes over, just as above.
Second, the leading IR renormalon contribution will no longer cancel completely
as it does in eq.~\eqn{cgasymp}. However, it is true in general that it is
suppressed by a factor $1/n^2$ in the sum $c_{n,1}+g_n$ relative to $c_{n,1}$
alone \cite{benPhD}, since this follows from the OPE and the anomalous
dimension of the gluon condensate. Thus, we expect some of the features of the
large-$\beta_0$ toy model to survive in the realistic QCD case. 

Next, we corroborate these findings with two further simplistic models, which
will also uncover another crucial ingredient for the building of a more
realistic model for higher order coefficients.

\subsection{Single pole models}

In order to separate the effects of a specific renormalon pole, and to give
support to our previous findings, we investigate a single IR renormalon pole
at position $p$, for which we shall consider the particular cases $p=2$ and
$p=3$. Explicitly, our single pole model for the Borel transform of the Adler
function series takes the form:
\begin{equation}
\label{BRIRp}
B[\wh D_p](u) \,\equiv\, \frac{d_p^{\rm IR}}{(p-u)} +
\sum\limits_{i=0}^3 d_i^{\rm PO} u^i \,.
\end{equation}
We fix the residue $d_p^{\rm IR}$ such that the perturbative coefficient
$c_{5,1}=283$, and the polynomial terms $d_i^{\rm PO}$ are adjusted such as
to reproduce the lower order coefficients according to eqs.~\eqn{cn1} and
\eqn{c41}. The reason behind employing $c_{5,1}$ to fix the residue is that we
like to work with a perturbative order at which a dominance of the leading IR
renormalon to the coefficient is expected, but this assumption is not crucial
for our argument.

\FIGURE[ht]{\includegraphics[angle=0, width=14cm]{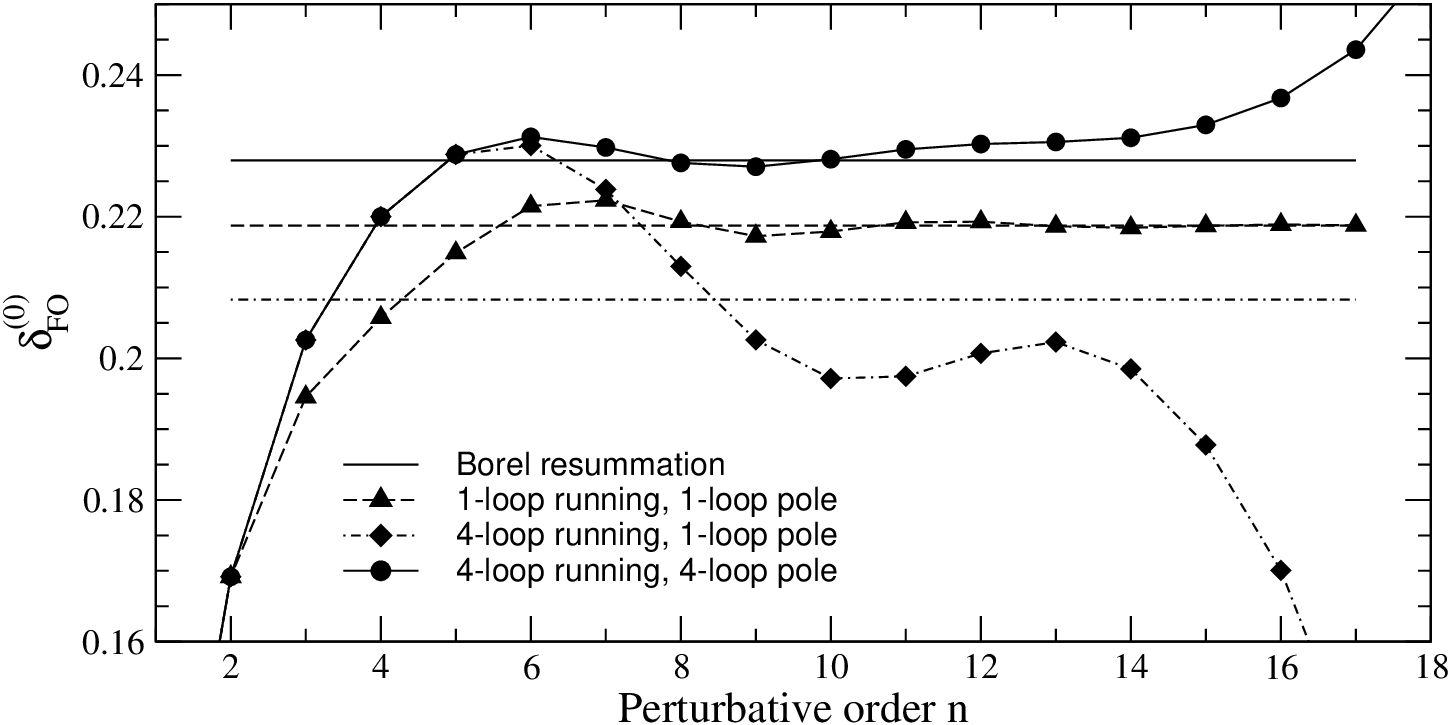}
            \includegraphics[angle=0, width=14cm]{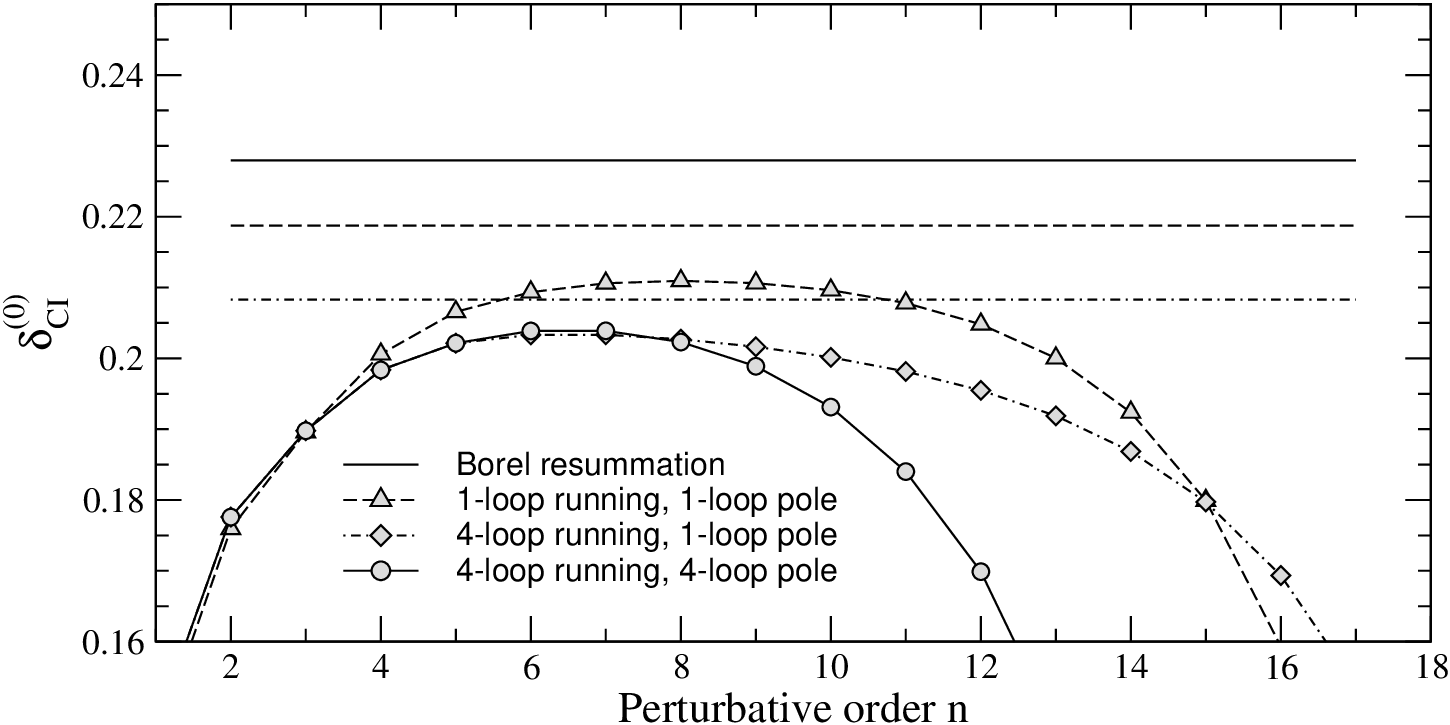}
\caption{Results for $\delta^{(0)}_{\rm FO}$ (full circles, diamonds,
triangles) and $\delta^{(0)}_{\rm CI}$ (grey circles, diamonds, triangles)
at $\as(M_\tau)=0.34$, for the model of eq.~\eqn{BRIRp} at $p=2$. The
triangles correspond to a running of $\as$ at 1-loop, while for the diamonds
a 4-loop running coupling has been employed. For the circles the pole in
eq.~\eqn{BRIRp} is modified to account for the 4-loop renormalon cut structure.
The straight dashed, dash-dotted and solid lines represent the respective Borel
sums.\label{fig4}}}

The numerical results for the model \eqn{BRIRp} with $p=2$ are shown in
figure~\ref{fig4}. The upper plot corresponds to $\delta^{(0)}_{\rm FO}$, the
lower one to $\delta^{(0)}_{\rm CI}$. Let us first concentrate on the triangles
which are connected by dashed line segments, and for which the running of $\as$
at 1-loop has been employed. This case is rather similar to the large-$\beta_0$
approximation. FOPT approaches the Borel sum (the horizontal dashed line) well,
although here in a slightly oscillatory manner, and up to the order shown in
the plot the divergent behaviour has not yet set in. Conversely, like for the
large-$\beta_0$ approximation above, for CIPT the divergent behaviour sets in
much earlier and the series never comes close to the Borel sum.

The picture drastically changes when the running of the coupling in the contour
integration is implemented at four loops. This case is shown as the full and
grey diamonds in figure~\ref{fig4}, which are connected by dash-dotted lines. 
The corresponding Borel sum is the straight dash-dotted line.
$\delta^{(0)}_{\rm FO}$ first overshoots the ``true'' result by a large amount,
then displays a large oscillation similar to figure~\ref{fig2}, before it
starts to diverge, while CIPT appears more like the 1-loop case. Still, with
4-loop running, neither FOPT nor CIPT approach the Borel sum in a sensible
fashion. The reason for this unexpected behaviour can be traced back to the
fact that our model \eqn{BRIRp} only contains a simple pole. Such a simple pole
is the correct structure of a renormalon pole in the large-$\beta_0$ limit, but
when higher terms in the $\beta$-function are to be included, the renormalon
pole structure gets more complicated, also involving cuts, whose gross features
are determined by the OPE~\cite{ben98}. Hence, to construct a consistent model
which aims to use a 4-loop running coupling, also the renormalon cut structure
has to be incorporated at the same order. The results required at four loops 
will be derived in the next section. Once four-loop running is consistently
included in the Adler function Borel transform eq.~\eqn{BRIRp} and the contour
integration, we obtain the circles connected by solid lines in
figure~\ref{fig4}. Now we find again that FOPT for an IR pole at $u=2$ smoothly
approaches the Borel sum, while CIPT fails. 

\FIGURE[ht]{\includegraphics[angle=0, width=14cm]{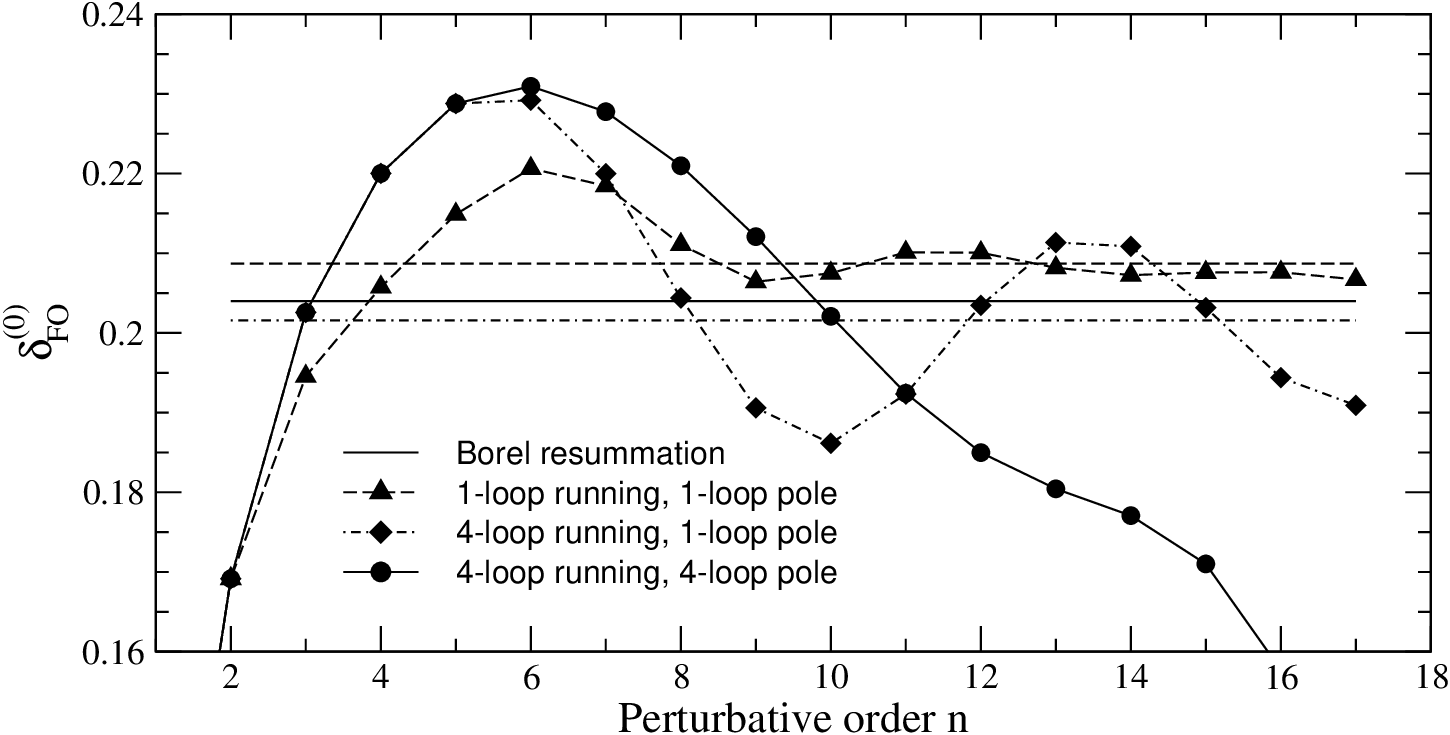}
            \includegraphics[angle=0, width=14cm]{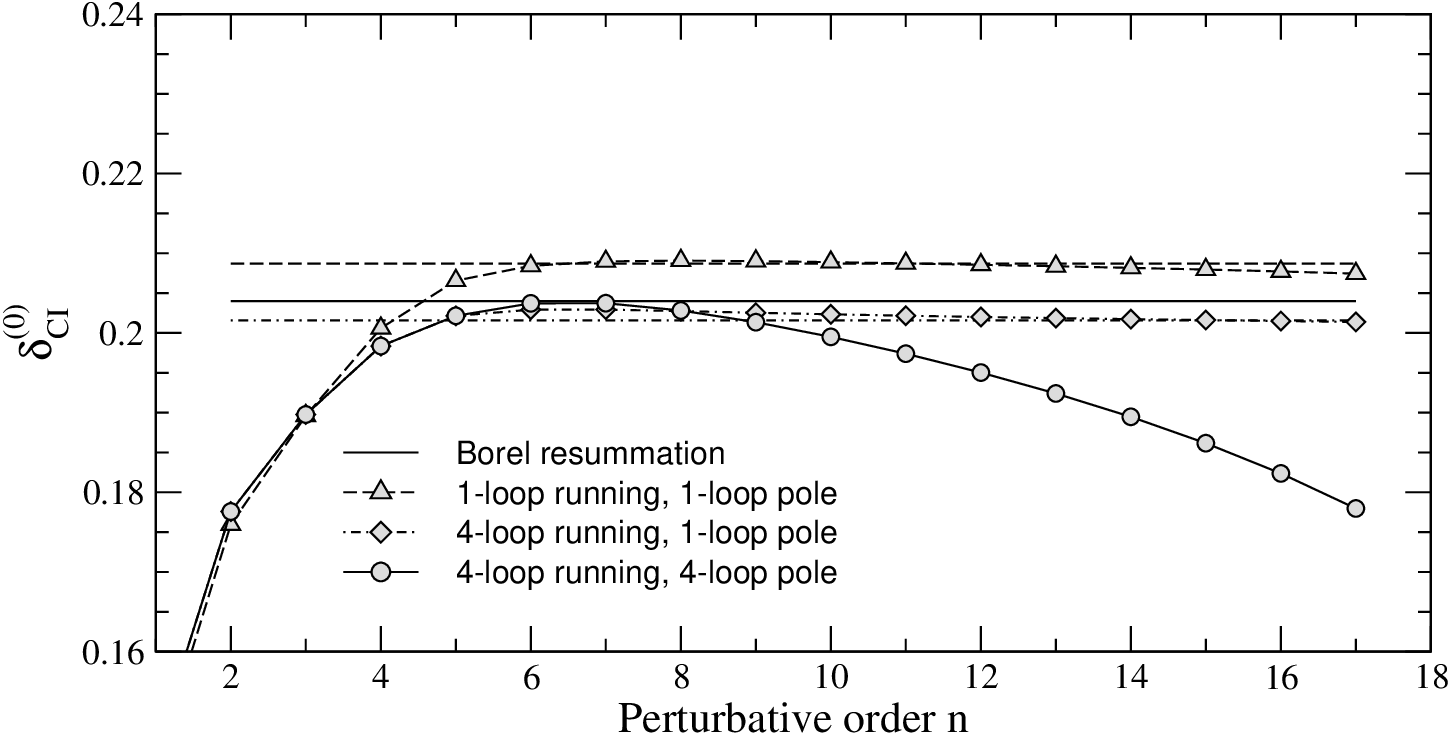}
\caption{Results for $\delta^{(0)}_{\rm FO}$ (full circles, diamonds, 
triangles) and $\delta^{(0)}_{\rm CI}$ (grey circles, diamonds, 
triangles) at $\as(M_\tau)=0.34$,
for the model of eq.~\eqn{BRIRp} at $p=3$. The triangles correspond to a running
of $\as$ at 1-loop, while for the diamonds a 4-loop running coupling has been
employed. For the circles the pole in eq.~\eqn{BRIRp} is modified to
account for the 4-loop renormalon cut structure. 
The straight dashed, dash-dotted and solid 
lines represent the respective Borel
sums.\label{fig5}}}

To conclude the discussion of simple toy models, we finally investigate the
ansatz \eqn{BRIRp} with $p=3$. Graphically, this case is displayed in figure
\ref{fig5}, with the same notation as used in figure~\ref{fig4}. We see that
this case very much resembles the model of figure~\ref{fig2}, where the higher
$c_{n,1}$ had been set to zero.  The $p=3$ model differs from the $p=2$ one
in two respects. First, it follows from eq.~\eqn{Brtau} that there is no
suppression of $c_{n,1}+g_n$ relative to $c_{n,1}$ in large orders. More
importantly, with $p=3$ the divergence of the series is milder. Hence, the
Adler function coefficients $c_{n,1}$ are much smaller than the running effects
$g_n$ along the complex contour, an expectation that can be verified by
explicitly investigating the $c_{n,1}$ in this model. Thus, like for the first
model, the truncated Adler function, CIPT provides a good account of the Borel
sum, and FOPT is only able to approach this value with large oscillations.

\subsection{Resum\'{e}}

The main conclusions from the toy examples are as follows. We find that CIPT
provides the better approximation whenever running coupling effects to the
series expansion of $R_\tau$ dominate over the intrinsic Adler function
coefficients as should have been expected. This is the case in
truncated perturbation theory and in models with weak factorial divergence,
such as the $p=3$ single-pole model. We find that FOPT provides the better
approximation, whenever there are systematic cancellations between the Adler
function and running coupling contributions. Such cancellations occur in the
$p=2$ single-pole model, the large-$\beta_0$ approximation, and in general
for the leading IR renormalon contribution to $R_\tau$. We also find that to
correctly account for these cancellations, the running coupling effects have
to be implemented at the same loop-order in the contour integral and the
renormalon structure of the Adler function Borel transform. The interesting
question is now which of these features is relevant to the ``real world''. 

The discussion of this section already allows to disfavour a large class of
models for the realistic Borel-transformed Adler function that were initially
considered by us. In this class fall models with renormalon poles that only
have integer power, because they fail to account consistently for running
coupling effects beyond 1-loop evolution. For the same reason -- besides not
using the available information on the known positions of the renormalon
poles -- also models based on Pad\'e approximation are of limited use. 

In order to be able to build a more realistic model for $B[\wh D](u)$, as a
prerequisite, employing the RGE and the structure of the OPE, in the next
section we shall derive the general form of the renormalon cut including
$\beta$-function effects up to 4-loop level.

\section{Renormalon poles at four loops}\label{sect5}

Our next aim is a physically motivated ansatz for the Borel transform of the
Adler function $B[\wh D](t)$, which incorporates all known exact results, and
on the basis of which we will be in a position to investigate the influence of
higher-order perturbative contributions, and to perform a comparison of the
perturbative series in FOPT and CIPT with its Borel sum. As has been seen in
the last section, if higher-order running effects are to be included in the
contour integration, the renormalon pole structure should match the
corresponding loop order. The derivation of the renormalon cut is based just
on the structure of the OPE and the RGE, detached from any limitations of the
large-$\beta_0$ approximation, up to an unknown overall constant. The
expressions obtained in this section extend the analysis already presented in
sections~3.2.3 and 3.3.1 of ref.~\cite{ben98} to one more loop order.

Let us begin with the IR renormalon poles. The central idea is that the IR
renormalon ambiguity of the Borel integral arises from long-distance regions in
Feynman integrals, and therefore must be consistent with the power-suppressed
terms appearing in the operator product expansion \cite{mue85,bb00}. Comparing
the energy dependence of a certain term in the OPE to the one of the complex
ambiguity of the Borel integral, the renormalon singularity that gives rise to
this ambiguity can be determined. A generic term in the OPE of $\wh D(s)$ from
an operator $O_d$ of dimension $d$ can be written as
\begin{equation}
\label{OPE}
\wh C_{O_d}(a_Q) \,\frac{\langle \wh O_d\rangle}{Q^d} \,=\,
[a_Q]^{\frac{\gamma_{O_d}^{(1)}}{\beta_1}} \left[\, \wh C_{O_d}^{(0)} +
\wh C_{O_d}^{(1)}\,a_Q + \wh C_{O_d}^{(2)}\,a_Q^2 + \ldots \,\right]
\frac{\langle \wh O_d\rangle}{Q^d} \,,
\end{equation}
where the anomalous dimension $\gamma_{O_d}$ of the operator $O_d$ is defined
by
\begin{equation}
\label{Omu}
-\,\mu\,\frac{d}{d\mu}\,O_d(\mu) \,\equiv\, \gamma_{O_d}(a_\mu)\,
O_d(\mu) \,=\, \left[\, \gamma_{O_d}^{(1)}\,a_\mu +
\gamma_{O_d}^{(2)}\,a_\mu^2 + \gamma_{O_d}^{(3)}\,a_\mu^3 + \ldots \,\right]
O_d(\mu) \,.
\end{equation}
For convenience, we have expressed eq.~\eqn{OPE} in terms of the scale
invariant operator $\wh O_d$, defined by
\begin{equation}
\label{Ohat}
\wh O_d \,\equiv\, O_d(\mu)\,\exp\biggl\{-\!\int \frac{\gamma_{O_d}(a_\mu)}
{\beta(a_\mu)}\,da_\mu \biggr\} \,,
\end{equation}
such that higher order coefficients of $\gamma_{O_d}$ are contained in the
Wilson coefficients $\wh C_{O_d}^{(k)}$. Since eq.~\eqn{OPE} will only be
needed up to a multiplicative factor, we do not have to specify the constant of
integration in \eqn{Ohat}, and without loss of generality, it can be assumed
to be zero.  Employing the RGE for $a_Q$, the $Q$-dependent part of \eqn{OPE}
can be written as:
\begin{eqnarray}
\frac{\wh C_{O_d}(a_Q)}{Q^d} &\,=\,& 
\mbox{const.}\times\,\wh C_{O_d}(a_Q)\,
{\rm e}^{-\frac{d}{\beta_1 a_Q}}
\left[ a_Q \right]^{-d\frac{\beta_2}{\beta_1^2}}
\exp\Biggl\{\,d\!\int\limits_0^{a_Q}\,\Biggl[\, \frac{1}{\beta(a)} -
\frac{1}{\beta_1 a^2} + \frac{\beta_2}{\beta_1^2 a} \,\Biggr] {\rm d}a
\Biggr\} \,, \nonumber \\
\label{CdoQd}
&\,=\,& \mbox{const.}\times \,\wh C_{O_d}(a_Q)\,
{\rm e}^{-\frac{d}{\beta_1 a_Q}}
\left[ a_Q \right]^{-d\frac{\beta_2}{\beta_1^2}}
\Big[\, 1 + b_1\,a_Q + b_2\,a_Q^2 + \ldots \,\Big] \,,
\end{eqnarray}
where the coefficients $b_1$ and $b_2$ are found to be:
\begin{equation}
b_1 \,=\, \frac{d}{\beta_1^3}\left( \beta_2^2 - \beta_1\beta_3 \right) \,,
\qquad
b_2 \,=\, \frac{b_1^2}{2} - \frac{d}{2\beta_1^4}\left( \beta_2^3 -
2\beta_1\beta_2\beta_3 + \beta_1^2\beta_4 \right) \,.
\end{equation}

To find the Borel transform that matches the $Q$-dependence of~\eqn{CdoQd},
we take the ansatz:
\begin{equation}
\label{BR3PIR}
B[\wh D_p^{\rm IR}](u) \,\equiv\, 
\frac{d_p^{\rm IR}}{(p-u)^{1+\tilde\gamma}}\,
\Big[\, 1 + \tilde b_1 (p-u) + \tilde b_2 (p-u)^2 +\ldots \,\Big] \,.
\end{equation}
Employing eq.~\eqn{ImRpIR}, the imaginary ambiguity corresponding to the Borel
integral of $B[\wh D_p^{\rm IR}](u)$ is found to be:
\begin{equation}
\label{ImR3P}
\IM\left[\wh D_p^{\rm IR}(a_Q) \right] \,=\, \mbox{const.}\times \,
{\rm e}^{-\frac{2p}{\beta_1 a_Q}}\,[a_Q]^{-\tilde\gamma}
\biggl[\, 1 + \tilde b_1\frac{\beta_1}{2} \,\tilde\gamma \,a_Q +
\tilde b_2\frac{\beta_1^2}{4} \,\tilde\gamma (\tilde\gamma-1)\,a_Q^2 
+\ldots \,\biggr] \,.
\end{equation}
Comparing eqs.~\eqn{CdoQd} and \eqn{ImR3P}, one deduces:
\begin{equation}
\label{gammat}
p \,=\, \frac{d}{2} \,, \quad
\tilde\gamma \,=\, 2p\,\frac{\beta_2}{\beta_1^2} -
\frac{\gamma_{O_d}^{(1)}}{\beta_1} \,, \quad
\tilde b_1 \,=\, \frac{2(b_1+c_1)}{\beta_1 \tilde\gamma} \,, \quad
\tilde b_2 \,=\, \frac{4(b_2+b_1 c_1+c_2)}
{\beta_1^2 \,\tilde\gamma(\tilde\gamma-1)} \,,
\end{equation}
where $c_1\equiv \wh C_{O_d}^{(1)}/\wh C_{O_d}^{(0)}$ and
$c_2\equiv \wh C_{O_d}^{(2)}/\wh C_{O_d}^{(0)}$. Taylor expanding the
ansatz~\eqn{BR3PIR} in $u$ and performing the Borel integral term by term
yields the perturbative series:
\begin{eqnarray}
\label{Ral3PIR}
\wh D_p^{\rm IR}(a_Q) &=& \frac{\pi d_p^{\rm IR}}
{p^{1+\tilde\gamma} \,\Gamma(1+\tilde\gamma)}\, \sum\limits_{n=0}^\infty\,
\Gamma(n+1+\tilde\gamma) \biggl(\frac{\beta_1}{2p}\biggr)^{\!n}
a_Q^{n+1} \nn \\
\smvs
&& \times \,\biggl[\, 1 + \frac{2 p}{\beta_1}
\frac{(b_1+c_1)}{(n+\tilde\gamma)} + \biggl(\frac{2 p}{\beta_1}\biggr)^{\!2}
\frac{(b_2+b_1 c_1+c_2)}{(n+\tilde\gamma)(n+\tilde\gamma-1)} +
{\cal O}\left(\frac{1}{n^3}\right) \,\biggr] \,.
\end{eqnarray}
Eq.~\eqn{Ral3PIR} extends the corresponding eq.~(3.51) of ref.~\cite{ben98}
to include terms of order $1/n^2$ in the large-order behaviour of the
perturbative series.\footnote{Note a missing sign in the global factor
containing $\beta_0$ in eq.~(3.51) of \cite{ben98}, which should read
$(-2\beta_0/d)^n$.}

The corresponding expression for a general UV renormalon pole when higher
orders in the running are included can be deduced by formally considering
negative coupling $a_Q$ and a Borel integral that ranges from zero to minus
infinity \cite{bbk97}. Then one finds poles leading to ambiguities on the
negative real axis, and the complex ambiguities can again be identified with
the RGE properties of some higher-dimensional operators. The result is that
the constants in the parametrisation of the general UV singularity,
\begin{equation}
\label{BRUV3P}
B[\wh D_p^{\rm UV}](u) \,\equiv\, \frac{d_p^{\rm UV}}{(p+u)^{1+\bar\gamma}}\,
\Big[\, 1 + \bar b_1 (p+u) + \bar b_2 (p+u)^2 \,\Big] \,,
\end{equation}
can be obtained from the corresponding parameters of the IR renormalon pole
\eqn{BR3PIR} with the replacement $p\to -p$, leading to:
\begin{equation}
\label{gammab}
\bar\gamma \,=\, -\,2p\,\frac{\beta_2}{\beta_1^2} +
\frac{\gamma_{O_d}^{(1)}}{\beta_1} \,, \qquad
\bar b_1 \,=\, -\,\tilde b_1[p\to -p] \,, \qquad
\bar b_2 \,=\,    \tilde b_2[p\to -p] \,.
\end{equation}
The perturbative expansion corresponding to eq.~\eqn{BRUV3P} takes the
form:\footnote{There is a sign mistake in the term proportional to $1/n$ in
the corresponding eq.~(3.48) of \cite{ben98}.}
\begin{eqnarray}
\label{RalUV3P}
\wh D_p^{\rm UV}(a_Q) &=& \frac{\pi d_p^{\rm UV}}
{p^{1+\bar\gamma} \,\Gamma(1+\bar\gamma)}\, \sum\limits_{n=0}^\infty\,
\Gamma(n+1+\bar\gamma) \biggl(-\frac{\beta_1}{2p}\biggr)^{\!n}
a_Q^{n+1} \nn \\
\smvs
&& \times \,\biggl[\, 1 +
\bar b_1\,\frac{p\,\bar\gamma}{(n+\bar\gamma)} + \bar b_2\,
\frac{p^2\,\bar\gamma(\bar\gamma-1)}{(n+\bar\gamma)(n+\bar\gamma-1)}
 + {\cal O}\left(\frac{1}{n^3}\right)\,\biggr] \,.
\end{eqnarray}

In the next section, the general forms of the IR and UV renormalon
singularities \eqn{BR3PIR} and \eqn{BRUV3P} will be employed to construct
a physically motivated model for the Borel transform of the Adler function
$B[\wh D](u)$.  In order to describe the leading IR renormalon at $u=2$ as
well as possible, in this case we include the known Wilson coefficient
function and anomalous dimension of the gluon condensate $\langle a G^2\rangle$
\cite{cgs85}, which results in \cite{ben93a}
\begin{equation}
c_1[\langle a G^2\rangle] \,=\, \frac{C_A}{2} - \frac{C_F}{4} -
\frac{\beta_2}{\beta_1} \,,
\end{equation}
while all other $c$'s appearing in the equations above will be set to zero.

\section{A physical model for the Adler function series}\label{sect6}

To clarify whether FOPT or CIPT results in a better approximation to the
$\tau$ hadronic width, we need to construct a physically motivated model for
the Adler function series beyond the order $n=4$, up to which it is known
exactly.  Since the four lowest coefficients are available, it is reasonable to
attempt to merge the low-order series with the expected large-order behaviour.
Furthermore, it is convenient to generate the series from its Borel
transform.\footnote{A previous attempt to resum the perturbative series for
$R_\tau$, based on Borel transforms of the Adler function, has been made in
ref.~\cite{cdls01}.} A physical model should account for the following
features:
\begin{itemize}
\item It should reproduce the exactly known $c_{n,1}$, $n\leq 4$.
\item The very large order behaviour is governed by a sign-alternating UV
  renormalon divergence. Since the low-order series shows no sign of a
  sizeable alternating component, it should be sufficient to include the
  leading singularity at $u=-1$. On the other hand, since the intermediate
  orders are governed by IR renormalons, at least two IR renormalon
  singularities, at $u=2$ and $u=3$, should be included to merge the
  large-order behaviour with the low-order exact coefficients.
\item Since four-loop running is employed in the contour integral that relates
  the Adler function to $R_\tau$, the renormalon singularities in the Borel
  transform cannot be simple poles, but should be extended consistently to
  cuts. This is particularly important for $u=2$. We use the equations from
  section~\ref{sect5}, incorporating $c_1[\langle a G^2\rangle]$ into the
  description of the $u=2$ cut.  For the UV renormalon singularity at $u=-1$,
  we use $\bar\gamma = 1-2\beta_2/\beta_1^2$ in eq.~\eqn{gammab}, since it is
  known to be a double pole in the large-$\beta_0$ approximation.\footnote{
  The additional ``1'' arises from the anomalous dimension term 
  $\gamma_{O_d}^{(1)}/\beta_1$ in eq.~\eqn{gammab} related to a 
  four-quark operator \cite{benPhD}.}
\end{itemize}
We are thus led to the ansatz: 
\begin{equation}
\label{BRu}
B[\wh D](u) \,=\, B[\wh D_1^{\rm UV}](u) + 
B[\wh D_2^{\rm IR}](u) + B[\wh D_3^{\rm IR}](u) +
d_0^{\rm PO} + d_1^{\rm PO} u \,,
\end{equation}
where the first three terms use \eqn{BR3PIR} and \eqn{BRUV3P} as building
blocks for the three leading renormalon singularities.\footnote{A similar
approach to the heavy quark mass has been put forward in
refs.~\cite{lee99,pin01}.} The model~\eqn{BRu} depends on five parameters,
the three residua of the renormalon poles, $d_1^{\rm UV}$, $d_2^{\rm IR}$ and
$d_3^{\rm IR}$, as well as the two polynomial parameters $d_0^{\rm PO}$ and
$d_1^{\rm PO}$. To fix these parameters, we match the perturbative expansion
of our model to the known coefficients $c_{n,1}$ of eqs.~\eqn{cn1}, \eqn{c41}
and \eqn{c51}. First, the residua are fixed using the coefficients $c_{3,1}$,
$c_{4,1}$ and $c_{5,1}$, and then the polynomial coefficients are adjusted to
also reproduce the first two orders $c_{1,1}$ and $c_{2,1}$. The motivation for
also making use of the fifth datum, namely the estimate for the coefficient
$c_{5,1}=283$, as well as including the polynomial terms, is that we wish to
avoid fixing the residue of a renormalon term by using orders as low as $n=2$.
However, as we discuss below, our result is surprisingly independent of this
extra assumption.

\TABLE[htb]{%
\renewcommand{\arraystretch}{1.1}
\begin{tabular}{rrrrrrr}
\hline\hline
$c_{6,1}\;$ & $c_{7,1}\quad$ & $c_{8,1}\quad$ & $c_{9,1}\quad$
& $c_{10,1}\quad$ & $c_{11,1}\quad$ & $c_{12,1}\quad$ \\
\hline
$3275$ & $1.88\cdot 10^4$ & $ 3.88\cdot 10^5$ & $ 9.19\cdot 10^5$ &
$ 8.37\cdot 10^7$ & $-5.19\cdot 10^8$ & $ 3.38\cdot 10^{10}$ \\
\hline\hline
\end{tabular}
\caption{Predictions for the Adler-function coefficients $c_{6,1}$ to
$c_{12,1}$ from our model \eqn{BRu} for $B[R](u)$, employing the estimate
for the coefficient $c_{5,1}=283$ of eq.~\eqn{c51}.\label{tab2}}}

Following the outlined procedure, the parameters of the model \eqn{BRu} are
found to be:
\begin{equation}
\label{dUVIR}
d_1^{\rm UV} \,=\, -\,1.56\cdot 10^{-2} \,, \qquad
d_2^{\rm IR} \,=\,    3.16  \,, \qquad
d_3^{\rm IR} \,=\, -\,13.5 \,, \\[-1mm]
\end{equation}
\begin{equation}
d_0^{\rm PO} \,=\,    0.781 \,, \qquad
d_1^{\rm PO} \,=\,    7.66\cdot 10^{-3} \,. \nn
\end{equation}
The fact that the parameter $d_1^{\rm PO}$ turns out to be so small implies
that the coefficient $c_{2,1}$ is already reasonably well described by the
renormalon pole contribution, although it was not used to fix the residua.
Another implication of this observation will be discussed below. With the
Borel transform thus determined, we are now in the position to investigate the
higher-order coefficients. The perturbative series of the Adler function up to
$c_{12,1}$ is presented in table~\ref{tab2}. We observe that the first negative
coefficient is found at the 11th order, after which the series retains its
sign-alternating behaviour. The successive approximations to the (reduced) 
Adler function $\wh D(M_\tau^2)$ defined in \eqn{Ralpha} are displayed in 
figure~\ref{figadler}, which shows that the model is well-behaved: the series
goes through a number of small terms at order $n=4$ to 7, the minimal term
being reached at $n=5$, such that the truncated series agrees nicely with its
Borel sum. The sign-alternating UV renormalon divergence takes over around
$n=10$.

\FIGURE[ht]{\includegraphics[angle=0, width=14cm]{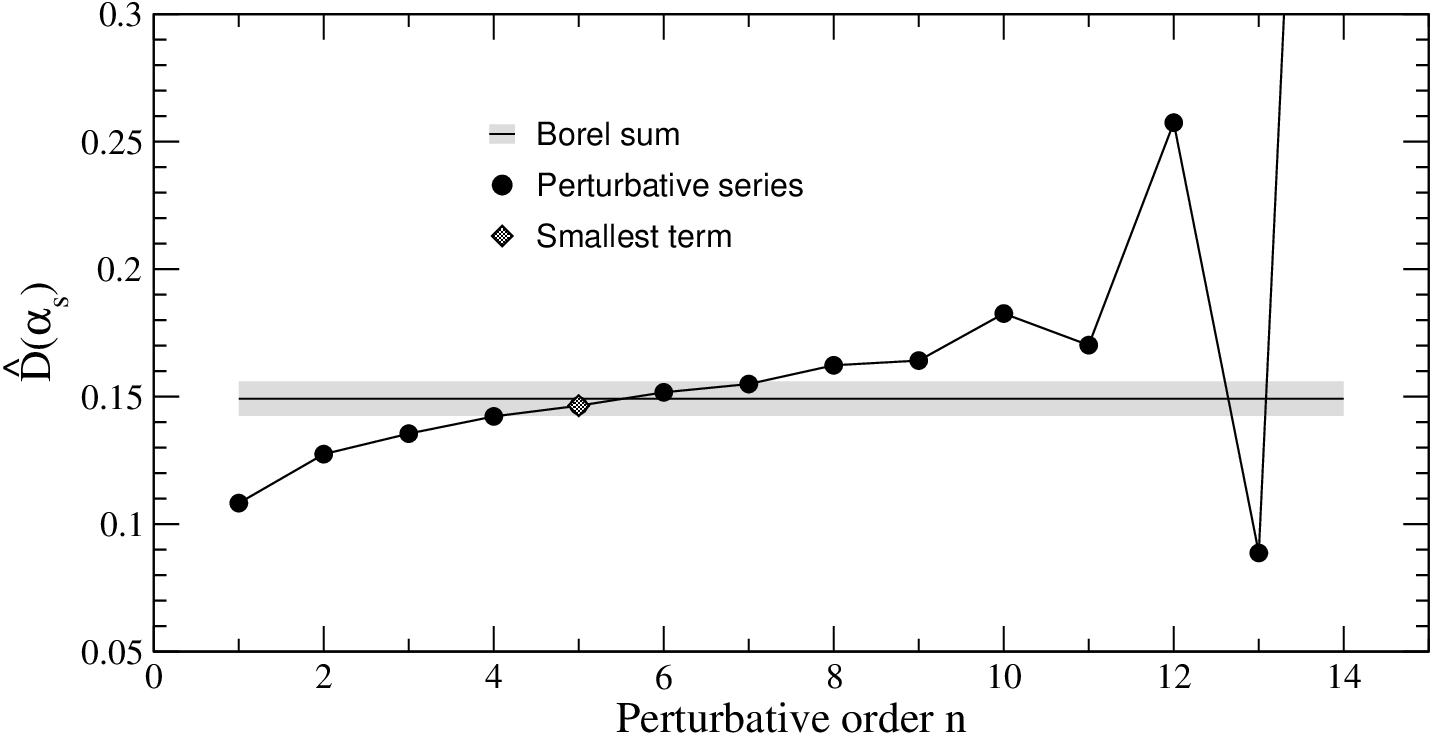}
\caption{Results for $\wh D(M_\tau^2)$ (full circles) at 
$\as(M_\tau)=0.34$, employing
the higher-order coefficients $c_{n,1}$ of table~\ref{tab2} obtained from our
model \eqn{BRu}, as a function of the order $n$ up to which the terms in the
perturbative series have been summed. The straight line represents the result
for the Borel sum of the series, and the shaded band provides an error estimate
inferred from the complex ambiguity.\label{figadler}}}

\TABLE[htb]{%
\begin{tabular}{rrrrrrrrrrr}
\hline\hline
& $c_{3,1}$ & $c_{4,1}$ & $c_{5,1}$ & $c_{6,1}$ & $c_{7,1}$ & $c_{8,1}$ &
  $c_{9,1}$ & $c_{10,1}$ & $c_{11,1}$ & $c_{12,1}$ \\
\hline
${\rm IR}_2$ & $82.4$ & $100.4$ & $135.9$ & $97.5$ & $155.9$ & $76.3$ &
               $359.9$ & $48.3$ & $-103.6$ & $22.9$ \\
${\rm IR}_3$ & $28.7$ & $-10.0$ & $-20.2$ & $-13.3$ & $-17.3$ & $-6.5$ &
               $-23.2$ & $-2.3$ & $3.6$ & $-0.6$ \\
${\rm UV}_1$ & $-11.2$ & $9.7$ & $-15.6$ & $15.8$ & $-38.6$ & $30.3$ &
               $-236.7$ & $54.0$ & $200.0$ & $77.7$ \\
\hline\hline
\end{tabular}
\caption{Relative contributions (in \%) of the different IR and UV renormalon
poles to the Adler-function coefficients $c_{3,1}$ to $c_{12,1}$.\label{tab3}}}

To gain further insight into the contribution of a certain renormalon
singularity to the coefficients $c_{n,1}$, in table~\ref{tab3} the relative
contributions (in \%) for the two IR and the UV pole are displayed. As can be
seen from table \ref{tab3}, already for the third order there is a reasonable
dominance of the leading IR renormalon at $u=2$, and a sizeable but not too
large contribution of the second one at $u=3$. For the next few orders the
leading IR pole becomes even more dominant as it should. Then, after a region
of cancellations between the leading IR and UV poles, the leading UV
renormalon, which dictates the large order behaviour, takes over. Nevertheless,
already at the fifth order, the last we have included in fixing our parameters,
with $16\%$ there is a noticeable contribution from the UV renormalon, so that
we have sensitivity to this term in the construction of the model.

\FIGURE[ht]{\includegraphics[angle=0, width=14cm]{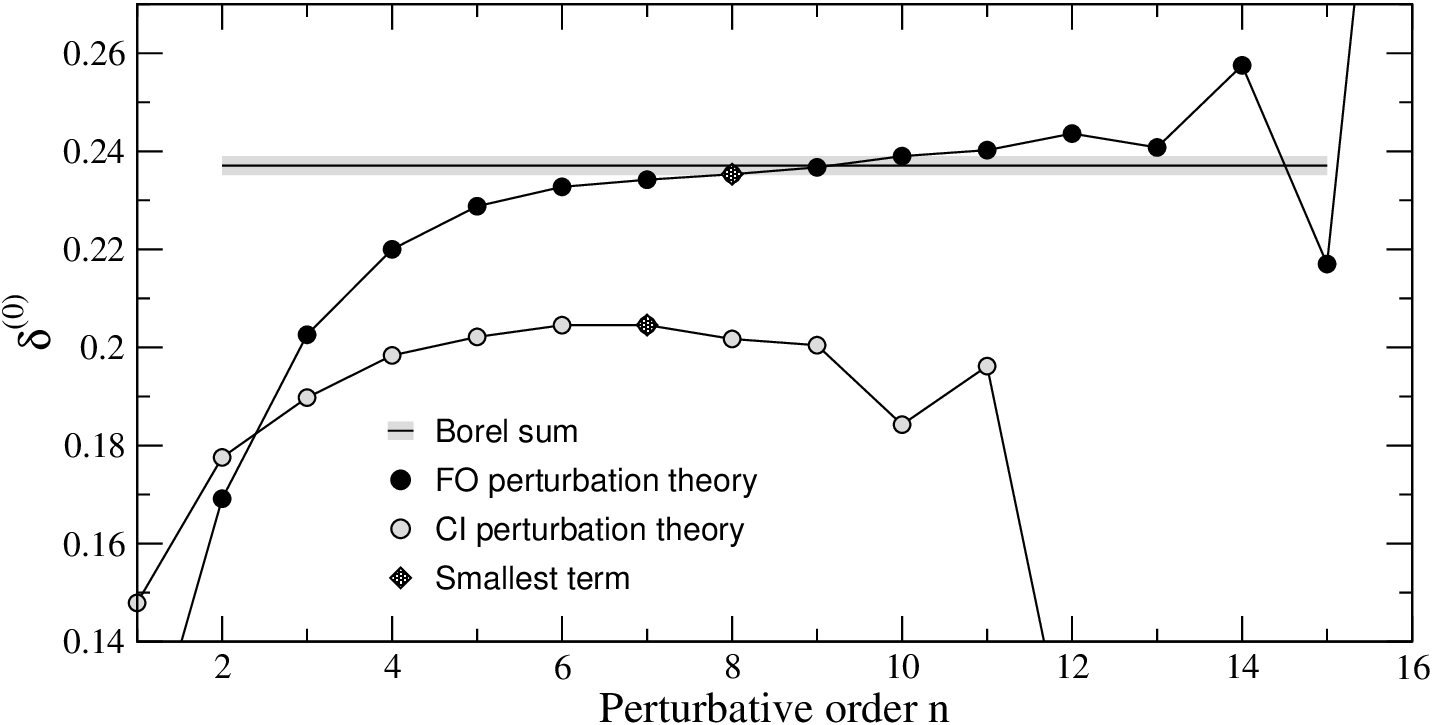}
\caption{Results for $\delta^{(0)}_{\rm FO}$ (full circles) and
$\delta^{(0)}_{\rm CI}$ (grey circles) at $\as(M_\tau)=0.34$, employing
the higher-order coefficients $c_{n,1}$ of table~\ref{tab2} obtained from our
model \eqn{BRu}, as a function of the order $n$ up to which the terms in the
perturbative series have been summed. The straight line represents the result
for the Borel sum of the series, and the shaded band provides an error estimate
inferred from the complex ambiguity.\label{fig6}}}

Let us now move to the implications of our model \eqn{BRu} for the $\tau$
hadronic width, that is for $\delta^{(0)}$ in FOPT and CIPT. A graphical
representation of the series behaviours is displayed in figure~\ref{fig6}.
Like in section~\ref{sect4}, the full circles denote our result for
$\delta^{(0)}_{\rm FO}$ and the grey circles the one for
$\delta^{(0)}_{\rm CI}$, as a function of the order $n$ up to which the
perturbative series has been summed. The straight line corresponds to the
principal value Borel sum of the series, and, like in figure~\ref{fig3} for
the large-$\beta_0$ approximation, the shaded band provides an error estimate
based on the imaginary part divided by $\pi$. The order at which the FO and
CI series have their smallest terms is indicated by the grey diamonds. For
CIPT this happens at the 7th and for FOPT at the 8th order. The essential
conclusion is that the qualitative behaviour of the realistic series is
determined by the features that were discussed in section~\ref{sect4} in the
context of the $p=2$ single-pole model. The cancellations between the Adler
function coefficients $c_{n,1}$ and the contributions from the contour
integral, $g_n$, soften the divergence of the $R_\tau$ series relative to the
Adler function series shown in figure~\ref{figadler}, allowing the FO series
to approach the ``true'' result around its minimal term, after which the
large-order asymptotic behaviour takes over. On the contrary, CIPT misses this
cancellation and always stays much below the ``true'' result. If CIPT were
used to determine $\as$ from the measured value of $R_\tau$, i.e.
$\delta^{(0)}$, a too large value of $\as$ would be extracted to
compensate for the deficit. We emphasise that the clear preference for FOPT 
holds even if we discard all higher-order terms and use only the series up
to $n=4$, as can be seen from figure~\ref{fig6}, provided the gross features
of our ansatz are correct.\footnote{Further insights into the origin of the
difference between $\delta^{(0)}_{\rm CI}$ and $\delta^{(0)}_{\rm BS}$ can
be found in appendix~B.}

Numerically, our central result for the Borel sum is found to be
\begin{equation}
\label{del0BS}
\delta^{(0)}_{\rm BS} \,=\, 0.2371 \pm 0.0060 \,i\,.
\end{equation}
Due to the suppression of the leading IR renormalon divergence by the contour
integration, only one third of the imaginary part arises from the first IR pole,
and two thirds from the second. This is consistent with the power-suppressed
terms in the OPE, where for $R_\tau$ the $1/M_\tau^6$ terms dominate over the
$1/M_\tau^4$ terms for the same reasons. The ambiguity that our procedure
assigns to the ``true'' value is also of order of the power corrections as
discussed in section~\ref{sect7}.  At the order of its minimal term, the FOPT
result reads $\delta^{(0)}_{\rm FO}=0.2353$ with a minimal term of $0.0011$.
Thus, the difference of the Borel sum and the supposedly best result for
$\delta^{(0)}_{\rm FO}$ is about $1.6$ times the minimal term, and the
imaginary ambiguity divided by $\pi$ precisely is able to account for this
difference. Thus, our ansatz passes the requirements for a sensible asymptotic
series and OPE expansion.  In the remainder of this section, we shall make
further checks as to the robustness of the prediction for $\delta^{(0)}$ from
our model for $B[\wh D](u)$, and we estimate the uncertainty of our value for
$\delta^{(0)}_{\rm BS}$ given in \eqn{del0BS}.

Our first test consists in dropping $c_{5,1}$ as an input, and instead using
$c_{2,1}$ to determine the residua of the renormalon poles. In this case we
set $d_1^{\rm PO}=0$, since we have only four data to fix the parameters. 
The visual appearance of the corresponding plot for $\delta^{(0)}$ is 
indistinguishable from figure~\ref{fig6}, so we do not show it again. Also
numerically, with $\delta^{(0)}_{\rm BS}=0.2370 \pm 0.0061\,i$, we have
practically the same result as in our main model. However, now we are in a
position to predict the coefficient $c_{5,1}$, with the surprising result
$c_{5,1}=280$, almost the same as our naive estimate \eqn{c51} that was used
as an input before. This outcome was already anticipated by the fact that for
our main model \eqn{BRu}, $d_1^{\rm PO}$ (which we now set to 0) turned out to
be very small. However, when one breaks down $c_{2,1}$ into the contributions
from the different renormalon terms, one finds that the IR pole at $u=3$ is
twice as important as the leading IR pole, so the agreement with the previous
ansatz might be fortuitous. For this reason, we prefer to base our discussion
on an estimate of $c_{5,1}$, despite the obvious drawback of having to rely
on an additional assumption.

Another modification to corroborate our assumptions and the findings for the
main model is to add a third IR renormalon pole at $u=4$, again only adding
one constant $d_0^{\rm PO}$, and fitting the parameters to the first five
perturbative coefficients. Hence, like in the last model, the coefficient
$c_{2,1}$ influences the residua of the renormalon poles. The numerical
outcome is $\delta^{(0)}_{\rm BS}=0.2377 \pm 0.0065\,i$, once more only a
slight shift compared to \eqn{del0BS}. The small change in the complex
ambiguity only originates from the first two IR poles with almost no
contribution from the third. Also graphically, no difference would be visible
in comparison to figure~\ref{fig6}. When inspecting the separate pole
contributions, the third IR pole mainly contributes to the first and second
coefficient. For $c_{3,1}$, we are left with a small $5\%$ contribution, and
beyond the third order, the IR pole at $u=4$ is completely negligible.
Thus, our assumptions on the relevant singularities appear to be justified.

Two further sources of uncertainty in our result \eqn{del0BS} arise from the
dependence on the estimate of the coefficient $c_{5,1}$, which is not exactly
known, and from neglecting higher order running effects beyond $\beta_4$,
where the $\beta$-function coefficients are not known. In view of the fact
that $c_{5,1}$ turned out extremely stable when it was predicted above from
our model, to estimate the corresponding uncertainty, we think it is a save
choice to vary $c_{5,1}$ by $\pm\,50\%$. This results in variations of
$(\pm\,0.0042,\,\pm\,0.0030)$ in the real and imaginary part of
$\delta^{(0)}_{\rm BS}$, respectively, with no qualitative change in the
behaviours of the FOPT and CIPT series relative to 
figure~\ref{fig6}.\footnote{This remains true for larger variations of
$c_{5,1}$. However, for $c_{5,1}=0$ or negative, large cancellations between
the two IR renormalon poles are required in our model to produce such small
$c_{5,1}$ and the assumption to fit the low orders with only a few renormalon
singularities becomes questionable.} To get an idea about the importance of
higher order running effects, we can set $\beta_4$, $\bar b_2$ and $\tilde b_2$
to zero, and check how much our result changes. This results in a
$(-\,0.0021,\,-\,0.0011)$ variation 
of $\delta^{(0)}_{\rm BS}$. Adding both variations in quadrature, we arrive
at our final result for $\delta^{(0)}_{\rm BS}$ including uncertainties,
\begin{equation}
\label{del0BSun}
\delta^{(0)}_{\rm BS} \,=\, ( 0.2371 \pm 0.0047 )  \pm
                            ( 0.0060 \pm 0.0032 )\, i \,,
\end{equation}
which will be used in the next section for our determination of $\as$ from
the hadronic $\tau$ decay width. This should be compared to the perturbative
corrections from eqs.~\eqn{del0FOn} and \eqn{del0CIn}, which give 0.2200
(0.2288) in FOPT and 0.1984 (0.2021) in CIPT when the series is truncated
at $n=4$ ($n=5$).

Note that there is no uncertainty due to scale- or scheme-dependence in the
usual sense, since we are considering a series to all orders that is formally
scale- and scheme-independent. A question that can be asked, however, is
whether our ansatz for the Borel transform would still be meaningful and lead
to the same results regarding the validity of FOPT and CIPT, if the input data
$c_{n,1}$, $n\leq 5$, were given in another scheme for $\as$ than the
$\MSb$ scheme or at another scale. This is not obvious, but it is not clear
what conclusions should be drawn from this. An arbitrary scheme change can
produce arbitrary irregularities in the input data, making any attempt to
merge low with high orders meaningless. Similarly, a scheme change defined by
the relation $1/\as=1/\as^{\MSb}+\beta_1 C/(2\pi)$ would produce an additional
factor $\exp(-C u)$ in the new Borel transform, which changes the renormalon
residues and therefore the balance between contributions from the different
leading singularities. With no guidance at hand, we adopt the point of view
that the $\MSb$ scheme has proven useful and stable in so many applications of
perturbative QCD that there is little motivation to consider significant
departures. Nevertheless, to acquire an estimate of the model dependence of
our approach, in the next section we also fit our model parameters to the Adler
function coefficients which correspond to the expansion in the strong coupling
evaluated at a scale $\as(\xi M_\tau)$, and determine the impact of this
variation on the $\as$ determination. With such variations, the strength
of the leading UV renormalon is modified as compared to the lowest IR
renormalon pole. For $\xi<1$ the leading UV renormalon becomes stronger,
leading to an earlier onset of the asymptotic regime, while for $\xi>1$ it is
suppressed, which imposes limitations on sensible values of $\xi$.

\section{Determination of \boldmath{$\as$}}\label{sect7}

The starting point for a determination of $\as$ from hadronic $\tau$ decays
is eq.~\eqn{RtauVA} for the decay rate of the $\tau$ lepton into light $u$ and
$d$ quarks. The general strategy for our extraction of $\as$ below will be as
follows. We concentrate on the sum of the vector and axialvector channels,
$R_{\tau,V+A}$, as in this case some of the higher-dimensional operator
contributions cancel. A more elaborate analysis of the separate channels along
the lines of refs.~\cite{aleph05,dhz05,ddhmz08} employing moments of the $\tau$
decay spectral function is left for the future. We expect significantly larger
perturbative corrections and ambiguities for the moments, along with the
enhanced condensate contributions, which makes the $\as$ analysis more
complicated. We base our analysis on FOPT, and FOPT together with the ansatz
for higher-order terms of section~\ref{sect6}, but do not use CIPT, despite
the fact that the CI perturbation series appears to be better behaved in low
orders. The reason is that the analysis in the previous sections shows that
the better stability and smaller scale dependence of CIPT reflects the plateau
of the CIPT curve in figure~\ref{fig6}, but the plateau value of the
perturbative correction is far from the true value. In such a situation, 
theoretical error estimates based on scale dependence provide a very large
underestimate of the true uncertainty.

The first step of the $\as$ analysis consists in estimating the values of
the power corrections $\delta_{ud,V+A}^{(D)}$ in eq.~\eqn{RtauVA}, which arise
from higher-dimensional operators in the framework of the OPE. The power
corrections will be calculated in FOPT.\footnote{In principle, for quantities
other than the purely perturbative corrections to the Adler function, the
preference of FOPT over CIPT has to be investigated anew. However, the power
corrections are already a small correction, and for our application the
difference in treating the perturbative expansion of the Wilson coefficients
is irrelevant.} Given these estimates and experimental data, we calculate a
phenomenological value of $\delta^{(0)}$ using eq.~\eqn{RtauVA}. We then
determine the value of $\as(M_\tau)$ by requiring that the theoretical value
$\delta^{(0)}_{\rm theo}$ in FOPT, and in our model of section~\ref{sect6}
matches the phenomenological value $\delta^{(0)}_{\rm phen}$. Errors are
estimated by varying all parameters within their uncertainties.

\subsection{Estimate of power corrections}

Before going through the estimation of the power corrections, let us remark
that these will only be considered for the transversal part of the correlator
$\Pi_{ud}^{V+A,(1+0)}(s)$. The longitudinal contribution $\delta_{ud,S+P}$ to
$R_{\tau,V+A}$, arising from $\Pi_{ud}^{V+A,(0)}(s)$ which is related to scalar
and pseudoscalar correlation functions, will be included later according to a
phenomenological approach which was already employed in the determination of
$|V_{us}|$ from $R_\tau$~\cite{gjpps03,gjpps04}. This approach has much smaller
uncertainties than making use of the corresponding QCD expressions for the
scalar/pseudoscalar light-quark correlators since their perturbative expansions
converge substantially slower than those for the vector/axialvector correlation
functions. With foresight, we shall use $\as(M_\tau) = 0.3156 \pm 0.006$
in the numerical estimate of the power correction below.

The lowest-dimensional power corrections to $R_{\tau,V+A}$ are of dimension-2
and only arise from terms proportional to quark masses squared. Detailed
expressions for these contributions were given in ref.~\cite{pp99,ck93} up
to order $\as^2$, and the next corrections at ${\cal O}(\as^3)$ were first
presented in ref.~\cite{bck04}. From these expressions we calculate the
corresponding $\delta^{(2)}_{m^2,V+A}$ in FOPT.  Contributions to this quantity
arise from two sources: one is proportional to $(m_u^2+m_d^2)$ and the other
to $m_s^2$. The second is due to internal strange-quark loops and only starts
at ${\cal O}(\as^2)$. While the perturbative series for the $(m_u^2+m_d^2)$
term displays a reasonable convergence, the $m_s^2$ one is very badly behaved,
such that in total the ${\cal O}(\as^3)$ term is about a factor of three times
the ${\cal O}(\as^2)$ term. To estimate the uncertainty in
$\delta^{(2)}_{m^2,V+A}$, we average the results either including or omitting
the ${\cal O}(\as^3)$ term, and take the spread as the error. For the quark
masses we use $m_u(M_\tau)=2.8\pm 0.5\,$MeV, $m_d(M_\tau)=5.0\pm 0.6\,$MeV
and $m_s(M_\tau)=97\pm 9\,$MeV, which derive from ref.~\cite{jop06}. Also
quadratically including the parametric uncertainties, which albeit play a
minor role, yields
\begin{equation}
\label{del2}
\delta^{(2)}_{m^2,V+A} \,=\, ( 3.1 \pm 8.6 )\cdot 10^{-5} \,.
\end{equation}
Although this contribution has a very large uncertainty, it will turn out to
be immaterial for the $\as$ determination.

Dimension-4 contributions arise from three possible sources: the gluon
condensate $\langle a G^2\rangle$, the quark condensate $\langle\bar qq\rangle$,
and $m_q^4$ corrections.  (See~e.g.~ref.~\cite{pp99} and references therein for
explicit expressions.) The quartic mass corrections are tiny, and thus we shall
drop them. Furthermore, being suppressed by $1/s^2$ because of the weight
function in $R_\tau$, the contour integral is only non-vanishing if there are
additional logarithms $\ln(-s)$, which first appear at order
$\as^2$.\footnote{This is yet another manifestation of the suppression of the
leading IR renormalon at $u=2$.} The explicit expressions for the two
contributions including the known terms are then found to be
\begin{eqnarray}
\label{delG2}
\delta^{(4)}_{\langle G^2\rangle,V+A} &=& \frac{11\pi^2}{4}\,a(M_\tau^2)^2
\,\frac{\langle a G^2\rangle}{M_\tau^4} \,, \\
\smvs
\label{delmqq}
\delta^{(4)}_{\langle\bar qq\rangle,V+A} &=& 54\,\pi^2\,
\frac{(m_u+m_d)\langle\bar qq\rangle}{M_\tau^4}\,\biggl[\, a(M_\tau^2)^2 +
\Big( \sfrac{517}{36} - \sfrac{8}{3}\zeta_3 + \Big( \sfrac{5}{6} -
\sfrac{4}{3}\zeta_3 \Big) \kappa R_s \Big) a^3 \,\biggr] \,,\;\;
\end{eqnarray}
where $\kappa\equiv\langle\bar ss\rangle/\langle\bar qq\rangle$,
$R_s\equiv 2m_s/(m_u+m_d)$, and we have assumed isospin symmetry for the up
and down quark condensates. Historically, the standard value for the gluon
condensate is $\langle a G^2\rangle=0.012\,\gev^4$ \cite{svz79}, and not much
progress has been made since then. Therefore, it will be employed as our
central value. As we only know the leading term for
$\delta^{(4)}_{\langle G^2\rangle,V+A}$, to be conservative, we assign a 100\%
uncertainty. For $\delta^{(4)}_{\langle\bar qq\rangle,V+A}$, the required quark
condensate can be calculated from the GMOR relation \cite{gmor68,jam02}, with
the result $\langle\bar qq\rangle(M_\tau)=-\,(272\pm 15\,\mev)^3$. With 40\%,
the next-to-leading order $\as^3$ correction is large, but still has
perturbative character. Thus, we include this term and take its size as an
estimate for the missing higher orders. Further employing
$\kappa=0.8\pm 0.3$ \cite{jam02}, and the light quark masses from above, we
obtain
\begin{equation}
\label{del4num}
\delta^{(4)}_{\langle G^2\rangle,V+A} \,=\, (3.3 \pm 3.3)\cdot 10^{-4} \,,
\qquad
\delta^{(4)}_{\langle\bar qq\rangle,V+A} \,=\, (-\,4.9 \pm 6.2)\cdot 10^{-5}\,.
\end{equation}
It may be remarked that the upper range of
$\delta^{(4)}_{\langle G^2\rangle,V+A}$ just corresponds to the complex
ambiguity of the $u=2$ renormalon pole divided by $\pi$ in our all-order
ansatz. It is gratifying to observe that they are of the same order of
magnitude, giving support to the size of this term. Nevertheless, also the
dimension-4 contributions  only play a minor role in the determination of
$\as$.

At dimension six, $\delta_{ud,V+A}^{(D)}$ receives contributions from the
three-gluon condensate $\langle g^3 G^3\rangle$, 4-quark operators, and
lower-dimensional operators times appropriate powers of quark masses. The
coefficient function of $\langle g^3 G^3\rangle$ vanishes at the leading order
and is therefore suppressed. Also the operators which get multiplied by quark
masses only give very small contributions. Thus we neglect these two types of
contributions and concentrate on the 4-quark condensates. At leading order,
one is confronted with three four-quark operators, while at higher orders
many more are generated \cite{bnp92,lsc86,ac94}. As it appears impossible to
determine all required condensates from phenomenology, the so-called
vacuum-saturation approximation (VSA) had been proposed \cite{svz79}. This
reduces all 4-quark condensates to squares of the quark condensate. With this
simplifying assumption, the leading dimension-6 contribution takes the form
\begin{equation}
\label{del6}
\delta_{\langle\bar qq\bar qq\rangle,V+A}^{(6)} \,=\, -\,\frac{512}{27}\,
\pi^3 \as\,\frac{\rho\langle\bar qq\rangle^2}{M_\tau^6} \,=\,
( -\, 4.8 \pm 2.9 )\cdot 10^{-3} \,,
\end{equation}
where we assume $\rho=2\pm 1$ for the numerical estimate. Our reasoning for
this is as follows: VSA is incompatible with the scale and scheme
dependence of the 4-quark operators \cite{ac94,jk86}. Therefore, it does not
make sense to include the next-to-leading order corrections. Conventionally,
one introduces the parameter $\rho$, which comprises the violation of the VSA.
A rough idea about the size of the parameter $\rho$ can be gleaned from
phenomenological fits of the dimension-6 contributions to $\Pi^V_{ud}$,
$\Pi^A_{ud}$, $\Pi^{V+A}_{ud}$, as well as $\Pi^{V-A}_{ud}$
\cite{aleph05,dhz05,ddhmz08,fgr04} (and references therein). Comparing the
phenomenological fits with the VSA, values for $\rho$ from one to about three
are found, which motivates the chosen range. At any rate,
$\delta_{\langle\bar qq\bar qq\rangle,V+A}^{(6)}$ constitutes the dominant
power correction.

We also include a crude estimate for the still higher-dimensional operator
corrections.  Not much is known about these contributions, apart from the
fact that they should be suppressed as compared to
$\delta_{\langle\bar qq\bar qq\rangle,V+A}^{(6)}$, since they carry at least
two more powers of $1/M_\tau$. Inspecting the fits of
refs.~\cite{aleph05,dhz05,ddhmz08} where also a dimension-8 contribution was
included, one infers that roughly it could be of order $10^{-3}$. Therefore,
we have added
\begin{equation}
\label{del8}
\delta_{ud,V+A}^{(8)} \,=\, ( 0 \pm 1 ) \cdot 10^{-3}
\end{equation}
in our total estimate of the uncertainty for power corrections. There are
potentially also (short-distance) instanton contributions to the $\tau$
hadronic width. The leading contribution of this type is a rapidly increasing
and uncertain function of $\as(M_\tau)$. For $\as(M_\tau) = 0.32$ it
has been estimated to contribute
$2\cdot 10^{-3}$ to $\delta^{(D)}_{ud,V+A}$ \cite{bbb93}. Nonetheless, we do
not include a further power correction uncertainty for this term.

To complete our summary of power corrections to $R_\tau$, we still have to
compute the longitudinal contributions which arise from scalar and pseudoscalar
correlators. Because the perturbative series for these correlators do not
converge very well, here we shall follow the approach of
refs.~\cite{gjpps03,gjpps04}. The contribution from the scalar correlator is
suppressed by a factor $(m_u-m_d)^2$, and it can be altogether neglected. The
main idea then is to replace the QCD expressions for the pseudoscalar
correlator by a phenomenological representation. The dominant contribution to
the pseudoscalar spectral function stems from the well known pion pole, giving
\begin{equation}
\label{delPSpi}
\delta_{ud,S+P}^\pi \,=\, -\,16\pi^2\,\frac{f_\pi^2 M_\pi^2}{M_\tau^4}
\Biggl(1-\frac{M_\pi^2}{M_\tau^2}\Biggr)^{\!2} \,,
\end{equation}
plus small corrections from higher-excited pionic resonances. Repeating the
analysis of section~3 of ref.~\cite{gjpps03} and updating the input parameters, we find
\begin{equation}
\label{delspin0}
\delta_{ud,S+P} \,=\, ( -\,2.64 \pm 0.05 )\cdot 10^{-3} \,.
\end{equation}
The uncertainty in \eqn{delspin0} has been estimated by setting the
higher-resonance contribution to zero, which clearly demonstrates that the
pion pole is dominant. Collecting all contributions, and adding the errors
in quadrature, we arrive at our total estimate of all power corrections:
\begin{equation}
\label{delph}
\delta_{\rm PC} \,=\, ( -\, 7.1 \pm 3.1 )\cdot 10^{-3} \,.
\end{equation}
Our value \eqn{delph} is consistent with the most recent fit to the $\tau$
spectral functions performed in ref.~\cite{ddhmz08}.

As a matter of principle, the OPE of the correlation functions in the complex
$s$ plane, even when integrated over a suitable energy interval \cite{pqw76},
could be inflicted with so-called ``duality violations'' \cite{shi00}. In our
case, these would come from the contour integral close to the physical region,
and even though suppressed by a double zero, could lead to additional
contributions or uncertainties. Within a model, these contributions were
recently investigated for hadronic $\tau$ decays \cite{cgp08}, and in
ref.~\cite{ddhmz08} the model was fitted to the $\tau$ spectral functions,
with the finding that possible additional contributions are below the $10^{-3}$
level. In view of these results, we shall omit possible duality violating
terms, but in future analyses which perform a simultaneous fit of higher order
OPE contributions, in analogy to \cite{ddhmz08}, it might be worthwhile to
include them.

\begin{boldmath}
\subsection{$\as$ analysis}
\end{boldmath}

Employing the value $R_{\tau,V+A}=3.479\pm 0.011$, which results from
eq.~\eqn{Rtauex} in conjunction with $R_{\tau,S}= 0.1615\pm 0.0040$
\cite{ddhmz08}, as well as $|V_{ud}|=0.97418\pm 0.00026$ \cite{th07}, from
eq.~\eqn{RtauVA} the phenomenological value for $\delta^{(0)}$ can be derived
after accounting for the electroweak corrections and subtracting the power
correction given in eq.~\eqn{delph}:
\begin{equation}
\label{de0ph}
\delta^{(0)}_{\rm phen} \,=\, 0.2042 \pm 0.0038_{\rm exp} \pm
0.0033_{\rm PC} \,=\, 0.2042 \pm 0.0050 \,.
\end{equation}
By far the dominant experimental uncertainty is due to $R_{\tau,V+A}$.
In the second error, we have also included the ones from $S_{\rm EW}$ and
$\delta_{\rm EW}'$ which should be considered theoretical (though they are not
power corrections). The final step in the extraction of $\as(M_\tau)$ now
consists in finding the values of $\as$ for which the phenomenological value
$\delta^{(0)}_{\rm phen}$ matches the theoretical prediction, either from the
finite order series, or from our model to all orders.

Let us begin by analysing the impact of using FOPT when the perturbative
expansion is employed up to and including the fifth order. Besides the exactly
known coefficients, we make use of our value \eqn{c51}, $c_{5,1} = 283$, and
to estimate the corresponding uncertainty, we have either removed or doubled
the fifth order term. A further theoretical error is added to account for the
residual renormalisation scale dependence of the fixed order series. We
estimate this by expressing the FO series in terms of $a(\mu^2)$ rather than
$a(M_\tau^2)$ and by varying $\mu$ between $1\,$GeV and $2.5\,$GeV. Going
through this procedure, and keeping the errors separate to clearly see the
importance of the various contributing uncertainties, we obtain:
\begin{eqnarray}
\label{astauFO}
\as(M_\tau) &=& 0.3203 \pm 0.0032_{\rm exp} \pm 0.0028_{\rm PC} \pm
0.0027_{c_{5,1}} \,{}^{+\,0.0105}_{-\,0.0052}\,\mbox{(scale)} \nn \\
\tvs
&=&
0.320 \pm 0.003_{\rm exp} \,{}^{+\,0.011}_{-\,0.006}\,\mbox{(th)} \,=\,
0.320^{\,+\,0.012}_{\,-\,0.007} \,.
\end{eqnarray}
Comparing our result \eqn{astauFO} with other determinations of $\as(M_\tau)$
from hadronic $\tau$ decays \cite{ddhmz08,bck08} that include the exact fourth
order term, we observe that our value is smaller by $0.012$ \cite{bck08} and 
$0.024$ \cite{ddhmz08}, up to twice the estimated error. This should not come
as a surprise since the analysis of ref.~\cite{ddhmz08} relied on the use of
CIPT, which increases $\as(M_\tau)$, see figure~\ref{fig6}, while in
ref.~\cite{bck08} an average of FOPT and CIPT is performed. The error in
eq.~\eqn{astauFO} is dominated by the scale error, which is significantly
larger than in CIPT.\footnote{The scale variation has a minimum at the scale
$\mu\approx 1.22\,$GeV, which might therefore be considered as an ``optimal''
scale. It is amusing to note that evaluating $\as$ at this scale, we obtain
$\as(M_\tau)=0.3151$, very close to the improved result \eqn{astau}.} 
However, as argued above, we find
that the small scale-dependence in CIPT is misleading when interpreted as a
measure of the theoretical uncertainty.

We evolve $\as(M_\tau)$ to the $Z$-boson mass scale, using the
$\beta$-function known to four loops \cite{rvl97,cza04}, and the matching
coefficients at flavour thresholds up to order $\as^3$
\cite{cks97}.\footnote{The evolution has been performed independently with
the Mathematica package RunDec \cite{cks00} and a private routine coded by
one of us (MJ), finding complete agreement.} The central values of the flavour
thresholds $\mu_c^*=3.729\,$GeV 
and $\mu_b^*=10.558\,$GeV are taken to correspond to the physical
thresholds of open $D$ and $B$ meson production. In the matching coefficients
also the charm and bottom quark masses are required, which we assume to be
$m_c(m_c)=1.28\pm 0.05\,$GeV and $m_b(m_b)= 4.20\pm 0.05\,$GeV \cite{qwg05} in
the $\MSb$-scheme. The uncertainties resulting from the evolution are estimated
as follows: like in section~\ref{sect6}, we compare to consistent 3-loop
running, i.e. setting $\beta_4=0$ and removing the ${\cal O}(\as^3)$
coefficient in the matching relation, and we vary the quark masses as well as
the flavour thresholds in the ranges $\mu_c^*/2<\mu_c<2\mu_c^*$ and
$\mu_b^*/2<\mu_b<2\mu_b^*$. Adding all uncertainties in quadrature, we arrive
at:
\begin{eqnarray}
\label{asMzFO}
\as(M_Z) &=& 0.1185 \pm 0.0004_{\rm exp} \,{}^{+\,0.0013}_{-\,0.0008}\,
(\mbox{th}) \pm 0.0002_{\rm evol} \nn \\
\tvs
&=& 0.1185^{\,+\,0.0014}_{\,-\,0.0009} \,.
\end{eqnarray}
Again, this value is lower than previous results from hadronic $\tau$ decays,
but it is in perfect agreement with the global averages of $\as(M_Z)$
\cite{pdg06,bet06}.

To further improve our determination of $\as$, we now include the 
higher order perturbative terms according to our model of
section~\ref{sect6}. To this end, we need to find the value of $\as$ such the
value of the Borel sum $\delta^{(0)}_{\rm BS}$ matches the phenomenological
value \eqn{de0ph}.\footnote{We do not include a separate error from the
ambiguity of $\delta^{(0)}_{\rm BS}$ (at $\as(M_\tau) = 0.3156$:
$\delta^{(0)}_{\rm BS} = 0.2042 \pm 0.0029 \,i$), since it is subsumed in the
error of $\delta_{\rm PC}$, which is several times larger.}
Again keeping the contributing uncertainties separate, we find:
\begin{eqnarray}
\label{astau}
\as(M_\tau) &=& 0.3156 \pm 0.0030_{\rm exp} \pm 0.0026_{\rm PC} \pm
0.0025_{c_{5,1}} \pm 0.0011_{\beta_4=0} \,{}^{+\,0.0034}_{-\,0.0029}\,
\mbox{(scale)} \nn \\
\tvs
&=& 0.3156 \pm 0.0030_{\rm exp} \pm 0.0051_{\rm th} \,=\, 0.3156 \pm 0.0059 \,.
\end{eqnarray}
Besides the uncertainties already discussed previously, in eq.~\eqn{astau} we
have also included a scale/model 
uncertainty according to the reasoning put forward
at the end of section~\ref{sect6}. As the scale variation, we chose
$0.5<\xi^2<1.5$, where $\mu\equiv\xi M_\tau$. (The corresponding scale interval
is $1.26\,\gev<\mu<2.18\,\gev$.) The employed range in $\xi$ is dictated by
the observation that for lower scales the contribution of the leading UV
renormalon is enhanced as compared to the leading IR renormalon pole, while
for larger scales it is suppressed. At the lower end for $\xi$, 
the contribution
of the leading UV renormalon to the coefficient $c_{3,1}$ is of a similar size
than the leading IR renormalon, while at the upper end, we are only left with
a mere $3\,\%$ contribution of the first UV renormalon to the coefficient
$c_{5,1}$, practically losing the sensitivity to this contribution. Therefore,
beyond the used range for $\xi$, our model ceases to make sense. 
For given $\xi$, we determine $\as(\xi M_\tau)$ such that the 
value \eqn{de0ph} is obtained and then evolve back to the scale 
$M_\tau$. In quoting
the final uncertainty of $\as(M_\tau)$, in \eqn{astau} we have used the larger
scale variation in order to have a symmetric final error. The
dependence of the successive perturbative approximations to $R_\tau$ 
on the choice of $\xi$ is shown in figure~\ref{fig8}, upper panel. 
The lower panel shows the
corresponding approximations in CIPT, which are again seen to lie below 
the ``true'' result for any reasonable value of $\xi$.

\FIGURE[ht]{\includegraphics[angle=0, width=14cm]{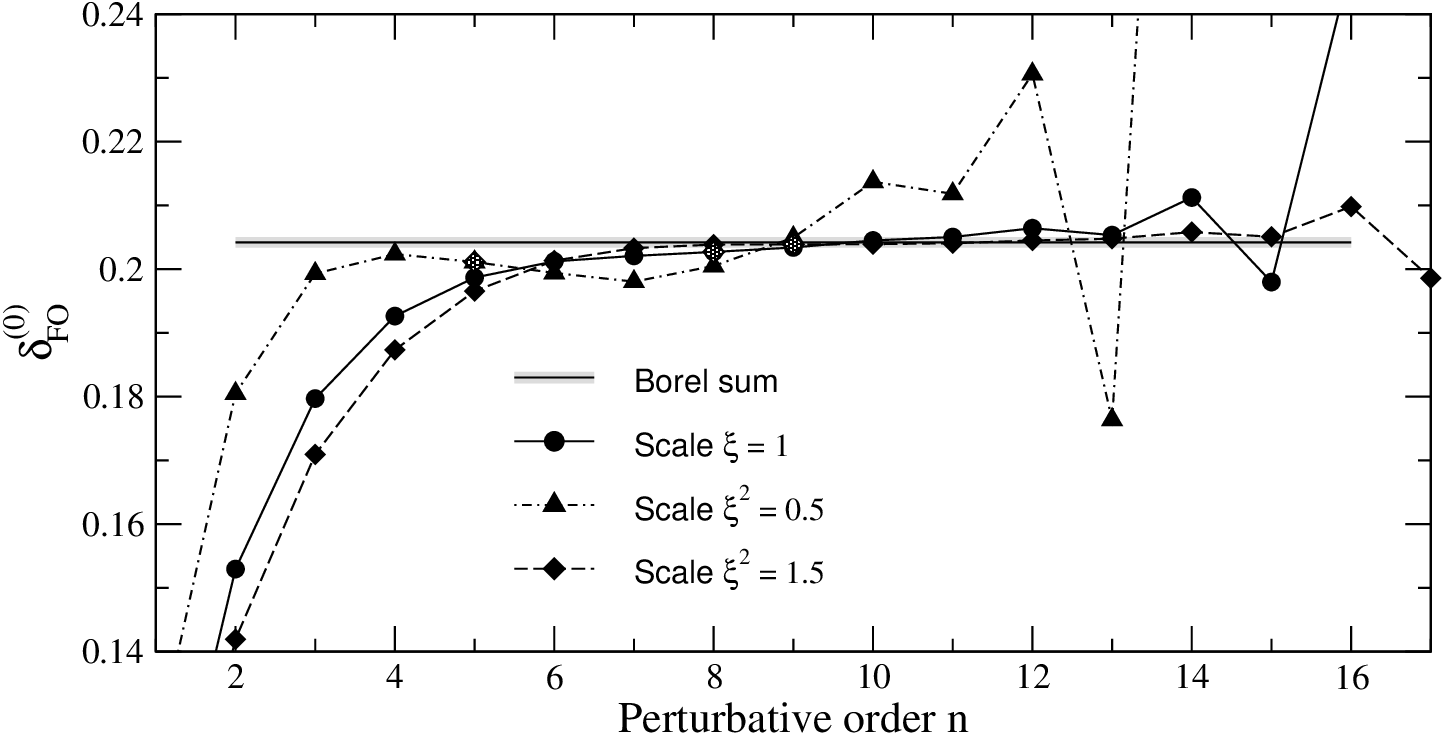}
            \includegraphics[angle=0, width=14cm]{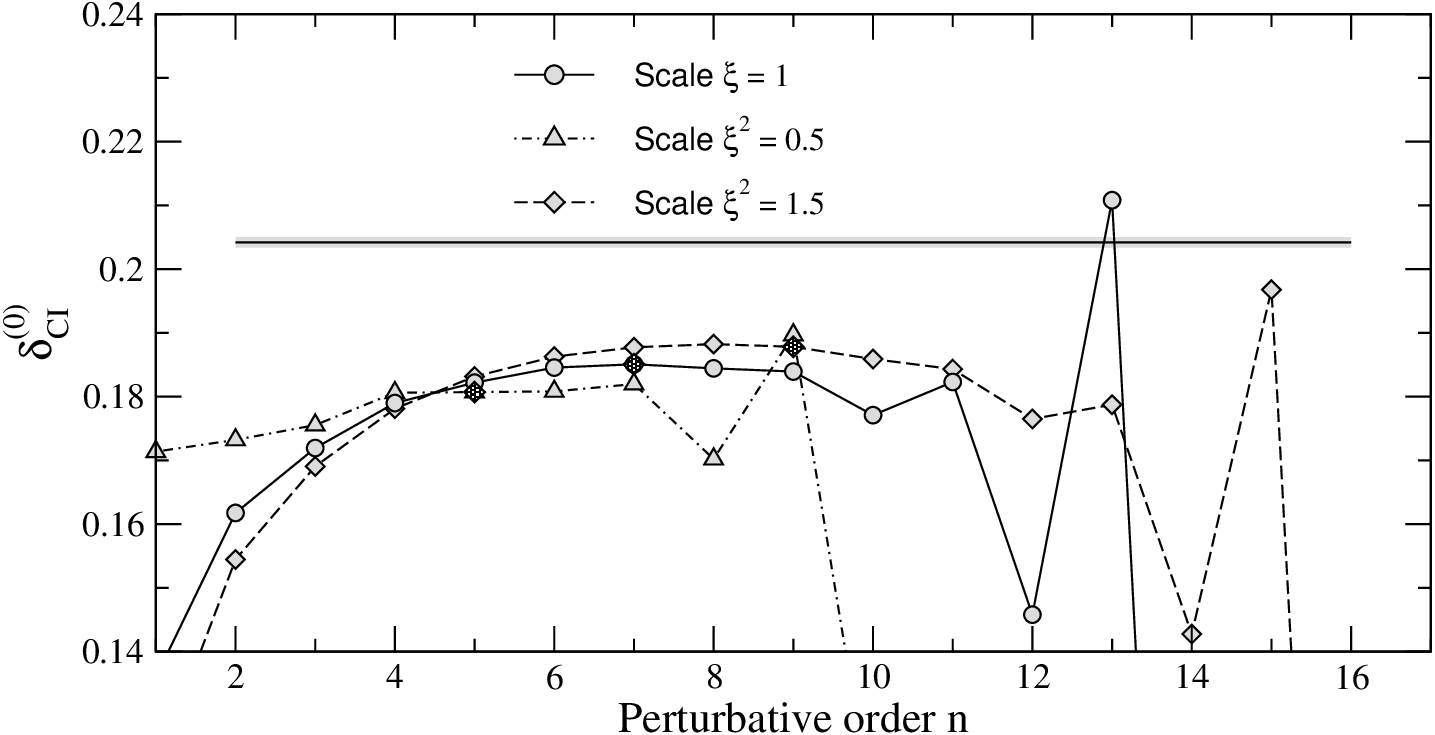}
\caption{Results for $\delta^{(0)}_{\rm FO}$ (full circles, diamonds, 
triangles) and $\delta^{(0)}_{\rm CI}$ (grey circles, diamonds, 
triangles) for the model of eq.~\eqn{BRu} as a function of the order $n$ 
up to which the terms in the perturbative series have been summed. 
The parameters of the model are fit to the first five coefficients 
of the Adler function expanded in $\as(\xi \sqrt{s})$ as described 
in section~\ref{sect6}. The value of $\as(\xi M_\tau)$ 
is determined by requiring that the Borel sum equals 
$\delta^{(0)}_{\rm phen} = 0.2042$ in each case.\label{fig8}}}

Evolving the result \eqn{astau} to the $Z$-boson mass scale, we arrive at our
final value for $\as(M_Z)$:
\begin{eqnarray}
\label{asMz}
\as(M_Z) &=& 0.11795 \pm 0.00038_{\rm exp} \pm 0.00063_{\rm th} \pm
0.00020_{\rm evol} \nn \\
\tvs
&=& 0.11795 \pm 0.00076 \,.
\end{eqnarray}
This result is slightly smaller than eq.~\eqn{asMzFO} based on the FOPT up to
the fifth order as could be anticipated from figure~\ref{fig6}, where one
observes that $\delta^{(0)}$ at ${\cal O}(\as^5)$ is smaller than the full
Borel sum. We consider eq.~(\ref{asMz}) as our best estimate of the strong
coupling in the $\overline{\rm MS}$ scheme from hadronic $\tau$ decays.

\section{Conclusions}\label{sect8}

Hadronic $\tau$ decays provide an especially clean environment for the study
of QCD effects and in particular the determination of QCD parameters. Of
prime interest in this respect is the QCD coupling $\as$. Still, due to the
relatively low scale $M_\tau$, an adequate control over the perturbative
series should be achieved for such applications. Besides explicitly computing
terms in the perturbative expansion, progress in this direction can be attained
by inspecting the renormalon divergence structure that the perturbative series
should have on general grounds.

A long-standing question in the interpretation of the QCD correction to
$R_\tau$ is a numerical discrepancy between two ways of performing the
renormalisation group improvement, namely CIPT and FOPT. While CIPT resums
running effects of $\as$ along the complex integration contour which are known
to be large, FOPT performs a consistent expansion in $\as$ at each loop order.
The CIPT sum is significantly below the FOPT sum requiring a larger strong
coupling to reproduce the experimental $R_\tau$. Resolving this discrepancy
has become a major issue for improving the accuracy of the $\as$
determination from $R_\tau$, in particular since recently the fourth order
coefficient of the series expansion has been computed \cite{bck08}, and further
terms cannot be expected to be available any time soon. 

CIPT has long been considered as the method of choice due to its apparent
better convergence and smaller renormalisation scale dependence of the
truncated series as compared to FOPT. However, the argument that FOPT should
be discarded, since the expansion of the running coupling along the contour
produces a series of finite radius of convergence that is avoided in CIPT,
does not take into account that both the CIPT and FOPT series have zero radius
of convergence anyway due to factorial divergence in higher orders. This leads
to interesting cancellations between the Adler function coefficients and 
running coupling effects \cite{bbb95}. Having a sufficiently large number of
exact Adler function coefficients at our disposal, and without prospects of
further improvements by exact computations, makes it timely and possible to
attempt to merge the exactly known terms with what is known on general grounds
about large-order behaviour, and to reinvestigate the conceptual issues of the
FOPT/CIPT comparison.

We investigated several toy models to find out under which circumstances FOPT
or CIPT provide a better approximation to the ``true'' result, which we may
define as the Borel sum of the series, since power corrections to $R_\tau$ are
small. Two extreme cases can be singled out. If the Adler function coefficients
are set to zero beyond a certain order, CIPT becomes exact, while FOPT performs 
large oscillations around the exact result~\cite{jam05}. On the other hand, if
the Adler function series is dominated by an infrared renormalon at $u=2$,
large cancellations occur in the $R_\tau$ series, which are only manifest in
FOPT. In this case, FOPT accounts well for the Borel sum, while CIPT stays
systematically below, despite an apparent better convergence and smaller scale
dependence! 

Taking these lessons we constructed a realistic ansatz for the entire Borel
transform of the Adler function such that the exactly known coefficients plus
an estimate of the fifth order coefficient are exactly reproduced, while
including knowledge about the leading three renormalon singularities. Given
the regularity of the exactly known low-order terms, the weight of the three
contributions was fit to the four- to six-loop terms. With our central model of
eq.~\eqn{BRu} we were able to achieve a very coherent picture of the known and 
higher order coefficients. Our main conclusion from a study of the realistic
model is that, given the particular features of the $R_\tau$ series, FOPT
provides the better approximation, in general and at fourth order, than CIPT 
and is to be preferred on these grounds. In this respect the full QCD case
resembles the large-$\beta_0$ approximation discussed in ref.~\cite{bbb95}.

We believe this conclusion to be valid in QCD beyond particular models, 
since it is based on generic properties of the $R_\tau$ perturbation 
series -- although it must be admitted that statements about 
the accuracy of FOPT and CIPT, and the validity of perturbation theory 
in higher orders in general, are always open to some amount of 
speculation. The main point is that conventional wisdom that favours 
CIPT is based on the first three orders in perturbation theory when 
the running coupling effects in $g_n$ dominate over the Adler function 
coefficients $c_{n,1}$. This dominance weakens as $n$ grows, as can be 
seen from the known exact coefficients, and the situation is expected 
to reverse beyond $n=5$. The fact that the $c_{n,1}$ ultimately 
diverge, cannot be ignored. This as well as the crucial $1/n^2$ 
cancellation in the sum $c_{n,1}+g_n$ for the leading non-sign-alternating 
component of the $R_\tau$ perturbation series are model-independent 
consequences of QCD.

Making use of these results, two determinations of the strong coupling $\as$
from $R_{\tau,V+A}$ were presented in section~\ref{sect7}. First, we performed
a conventional FOPT analysis based on the fourth-order result plus an estimate
of the fifth order, whose size is taken as an uncertainty. The resulting value
for $\as(M_\tau)$ can be found in eq.~\eqn{astauFO}. Evolving to $M_Z$, we
obtain 
\begin{eqnarray}
\as(M_Z) &=& 0.1185^{\,+\,0.0014}_{\,-\,0.0009} \qquad \mbox{(FOPT)}\,,
\end{eqnarray}
where the error is dominated by residual renormalisation scale dependence; 
see eq.~(\ref{asMzFO}). This value is lower than those presented in
refs.~\cite{ddhmz08,bck08} mainly because we propose to not use CIPT as the
result of our study. Our preferred and best result is 
obtained by incorporating 
available structural information on the large-order behaviour as embodied in
the model of eq.~\eqn{BRu}. The resulting value for $\as(M_\tau)$ can be found
in eq.~\eqn{astau}. Evolving it to $M_Z$, the main result of our paper reads:
\begin{eqnarray}
\as(M_Z) &=& 0.11795 \pm 0.00038_{\rm exp} \pm 0.00063_{\rm th} \pm
0.00020_{\rm evol} \nn \\
\tvs
&=& 0.11795 \pm 0.00076  \,.
\end{eqnarray}

Our approach towards the investigation of the higher-order behaviour of
perturbative series employed in this work can certainly also be applied to
other quantities of interest, like the scalar correlation function, mass
squared corrections to $R_\tau$, or moments of the spectral functions.
It should be interesting to see what can be said about the issue of
renormalisation group resummation when contour integrations are involved in
these cases. We shall return to this question in the near future. Finally,
our results can be useful in other places where the QCD Adler function plays
a role.

\bigskip
\acknowledgments
We would like to thank Santi Peris, Toni Pich, Antonio Pineda, Ximo Prades
and Felix Schwab for interesting discussions. We also thank our referee for
an interesting question which we answer in appendix~B. This work has been
supported in parts by the EU Contract No. MRTN-CT-2006-035482 (FLAVIAnet)
(MB, MJ), by CICYT-FEDER-FPA2005-02211 (MJ), by Spanish Consolider-Ingenio
2010 Programme CPAN (CSD2007-00042) (MJ), as well as by the DFG 
Sonderforschungsbereich/Transregio~9 ``Computer-gest\"utzte Theoretische
Teilchenphysik'' (MB).

\section*{Appendix A: Borel integral for renormalon poles}

\addcontentsline{toc}{section}
{Appendix A: Borel integral for renormalon poles}
\newcounter{alpha1} \renewcommand{\thesection}{\Alph{alpha1}} 
\renewcommand{\theequation}{\Alph{alpha1}.\arabic{equation}} 
\renewcommand{\thetable}{\Alph{alpha1}.\arabic{table}} 
\setcounter{alpha1}{1} \setcounter{equation}{0} \setcounter{table}{0}

In this appendix, we provide a few useful expressions for the Borel integral
of generic renormalon poles, and the corresponding complex ambiguity in the
case of IR renormalons. 

Let us begin with a general UV-renormalon pole which we assume to have the
form:
\begin{equation}
\label{BRUV}
B[R_p^{\rm UV}](u) \,\equiv\, \frac{d_p^{\rm UV}}{(p+u)^\gamma} \,,
\end{equation}
where $u=\beta_0 t$ with $\beta_0\equiv\beta_1/(2\pi)$, $p \in\mathbb N$ and
$\gamma\in\mathbb R^+$. Then, the corresponding Borel integral is given by
\begin{equation}
R_p^{\rm UV}(\alpha) \,=\, d_p^{\rm UV}\! \int\limits_0^\infty \!
\frac{{\rm e}^{-t/\alpha}}{(p+\beta_0 t)^\gamma}\,{\rm d}t \,,
\end{equation}
which after the substitution $t=\alpha z-p/\beta_0$ can be expressed in terms
of the incomplete $\Gamma$-function:
\begin{equation}
R_p^{\rm UV}(\alpha) \,=\, d_p^{\rm UV}\,\frac{\alpha}
{(\beta_0\alpha)^\gamma}\,{\rm e}^{p/(\beta_0\alpha)} \!\!\!\!
\int\limits_{p/(\beta_0\alpha)}^\infty \!\!\! z^{-\gamma}
{\rm e}^{-z}\,{\rm d}z \,=\, d_p^{\rm UV}\,
\frac{\alpha}{(\beta_0\alpha)^\gamma}\,{\rm e}^{p/(\beta_0\alpha)}\,
\Gamma\Big(1-\gamma,\sfrac{p}{\beta_0\alpha}\Big) \,.
\end{equation}
Expanding the UV-renormalon pole ansatz \eqn{BRUV}, and performing the Borel
integration term by term, yields the corresponding perturbative expansion:
\begin{equation}
\label{RUVal}
R_p^{\rm UV}(\alpha) \,=\, \frac{d_p^{\rm UV}}{p^\gamma \Gamma(\gamma)}\,
\sum\limits_{n=0}^\infty\,\Gamma(n+\gamma)
\biggl(\!-\,\frac{\beta_0}{p}\biggr)^{\!n} \alpha^{n+1} \,.
\end{equation}

For the general IR-renormalon pole, let us assume the generic form:
\begin{equation}
\label{BRIR}
B[R_p^{\rm IR}](u) \,\equiv\, \frac{d_p^{\rm IR}}{(p-u)^\gamma} \,,
\end{equation}
where again $u=\beta_0 t$, $p \in\mathbb N$ and $\gamma\in\mathbb R^+$. The
corresponding Borel integral is given by
\begin{equation}
R_p^{\rm IR}(\alpha) \,=\, d_p^{\rm IR}\! \int\limits_0^\infty \!
\frac{{\rm e}^{-t/\alpha}}{(p-\beta_0 t)^\gamma}\,{\rm d}t \,,
\end{equation}
and can be expressed in terms of the exponential integral function
${\rm E}_\gamma(z)$. However, because of the singularity at $t=p/\beta_0$
and the branch cut for $t>p/\beta_0$, we still have to specify the defining
integration path. We shall integrate on the real axis up to $t=p/\beta_0-\ve$,
then on a semicircle of radius $\ve$ in a clockwise, or anti-clockwise
direction, and finally again along the real axis from $t=p/\beta_0+\ve$ up to
infinity, either above or below the cut, taking the limit $\ve\to 0$ in the
end. Performing all integrals, this leads to the expression:
\begin{eqnarray}
\label{IRpIR}
R_p^{\rm IR}(\alpha) &=& d_p^{\rm IR}\,\frac{\alpha}{(\beta_0\alpha)^\gamma}
\,{\rm e}^{-p/(\beta_0\alpha)} \,\biggl\{\, -\,\big(
\sfrac{p}{\beta_0\alpha}\big)^{1-\gamma} \,{\rm E}_\gamma\big(-\sfrac{p}
{\beta_0\alpha}\big) \nn \\
\tvs
&& \hspace{37mm} +\,\left[\, (-1)^{\pm\gamma} -
(-1)^{{\rm sig}({\rm Im}[\alpha])\gamma} \,\right] \Gamma(1-\gamma)
\,\biggr\} \,.
\end{eqnarray}
In the above equation, $(-1)^z$ should be interpreted as $\exp(i\pi z)$, and
the function ${\rm sig}(z)$ represents the sign of $z$, with the additional
definition ${\rm sig}(0)\equiv 1$. The imaginary ambiguity of
$R_p^{\rm IR}(\alpha)$ can be readily computed from eq.~\eqn{IRpIR}, and
for $\alpha > 0$ turns out to be:
\begin{equation}
\label{ImRpIR}
\IM\left[ R_p^{\rm IR}(\alpha) \right] \,=\,
\pm\,\frac{d_p^{\rm IR}}{\beta_0^{\,\gamma}}\,\sin(\pi\gamma)\,
\Gamma(1-\gamma)\,\alpha^{1-\gamma}\,{\rm e}^{-p/(\beta_0\alpha)} \,.
\end{equation}
We can also straightforwardly obtain the perturbative expansion of
$R_p^{\rm IR}(\alpha)$, which reads:
\begin{equation}
\label{RIRal}
R_p^{\rm IR}(\alpha) \,=\, \frac{d_p^{\rm IR}}{p^\gamma \Gamma(\gamma)}\,
\sum\limits_{n=0}^\infty\,\Gamma(n+\gamma)\biggl(\frac{\beta_0}{p}\biggr)^{\!n}
\alpha^{n+1} \,.
\end{equation}

These results can be used to construct analytic expressions for our models,
which can all be decomposed into polynomials and a sum of poles of the above
form.

\begin{boldmath}
\section*{Appendix B: Adler function in the complex $s$-plane}
\end{boldmath}

\addcontentsline{toc}{section}
{Appendix B: Adler function in the complex \boldmath{$s$}-plane}
\newcounter{alpha} \renewcommand{\thesection}{\Alph{alpha}}
\renewcommand{\theequation}{\Alph{alpha}.\arabic{equation}}
\renewcommand{\thetable}{\Alph{alpha}.\arabic{table}}
\setcounter{alpha}{2} \setcounter{equation}{0} \setcounter{table}{0}

Further understanding of the origin for the difference of $\delta^{(0)}$ as
calculated within FOPT or CIPT can be gained by inspecting the reduced Adler
function $\wh D(s)$ on the circle $s=M_\tau^2\,{\rm e}^{i\varphi}$ in the
complex $s$-plane. From the general perturbative expansion \eqn{Ds} it is
clear that the ambiguity in the choice of the RG resummation already exists
in this case. Analogous to the discussion of section~\ref{sect3} for $R_\tau$,
FOPT is defined by employing the constant scale choice $\mu^2=M_\tau^2$,
while CIPT corresponds to a resummation of the series on each point of the
circle separately through the variable choice 
$\mu^2=-s=-M_\tau^2\,{\rm e}^{i\varphi}$. The perturbative expansions for
$\wh D$ as a function of the angle $\varphi$ in the two cases take the form:
\begin{eqnarray}
\label{DhatphiFO}
\wh D_{\rm FO}(\varphi) &=& \sum\limits_{n=1}^{\infty} 
a(M_\tau^2)^n \sum\limits_{k=1}^n k\,c_{n,k}\,[\,i(\varphi-\pi)]^{k-1} \,, \\
\smvs
\label{DhatphiCI}
\wh D_{\rm CI}(\varphi) &=& \sum\limits_{n=1}^{\infty} c_{n,1}\,
a(-M_\tau^2\,{\rm e}^{i\varphi})^n \,.
\end{eqnarray}

\FIGURE[ht]{\includegraphics[angle=0, width=14cm]{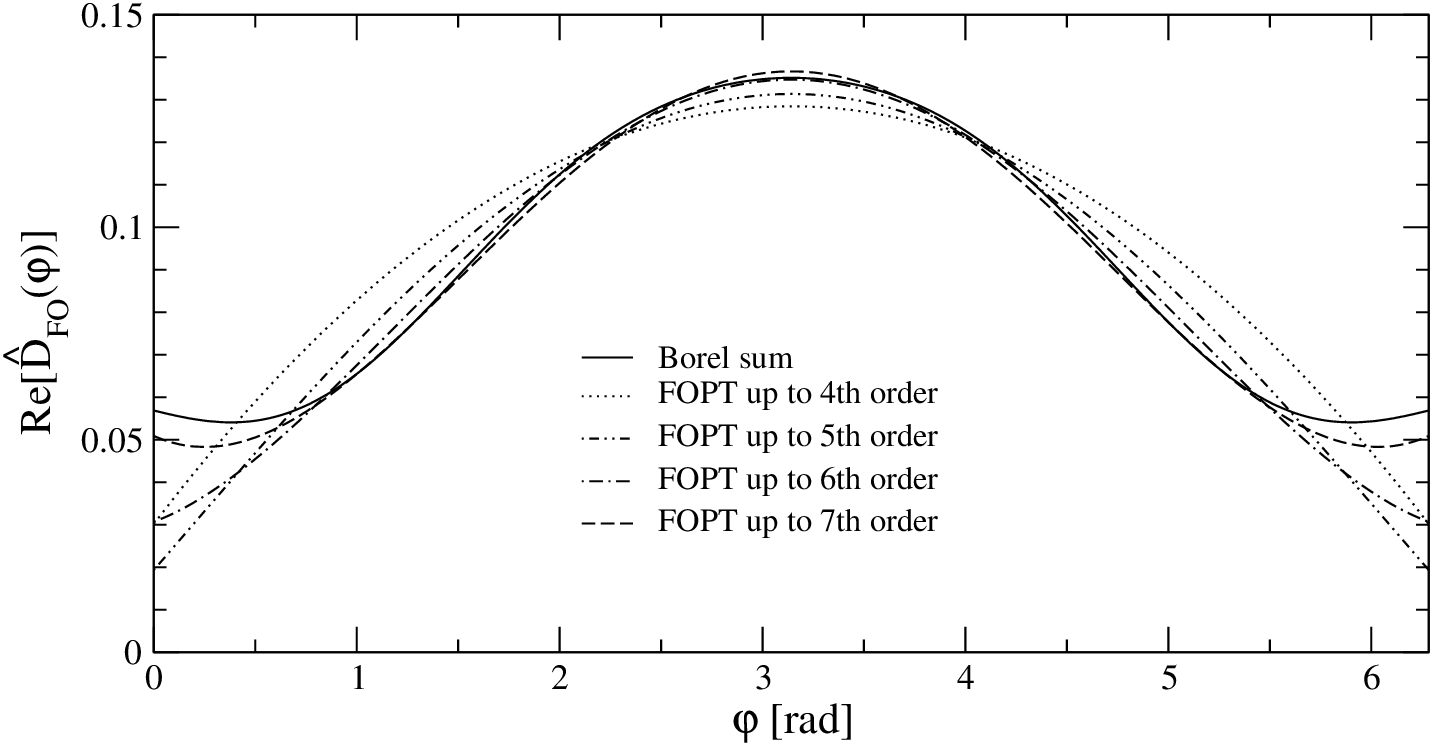}
            \includegraphics[angle=0, width=14cm]{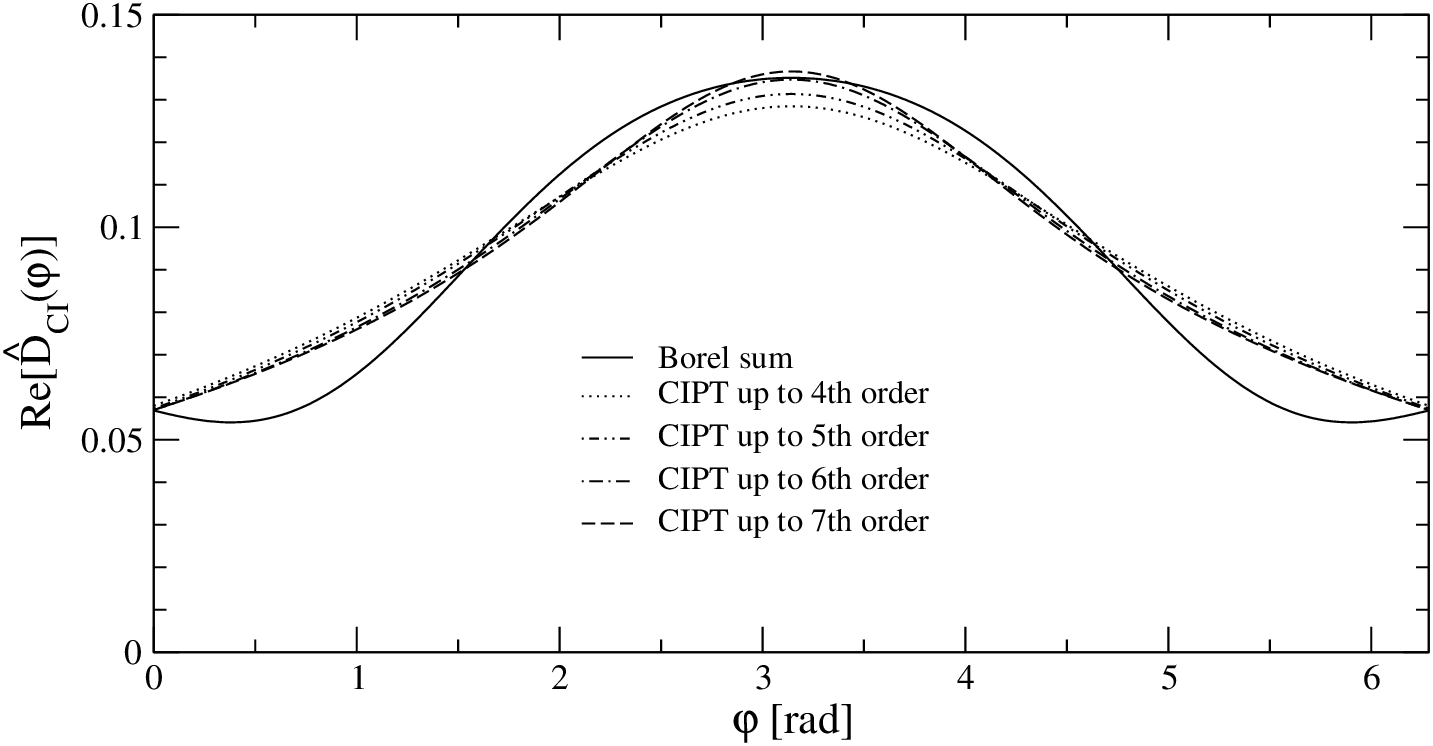}
\caption{The real parts $\RE[\wh D_{\rm FO}(\varphi)]$ and
$\RE[\wh D_{\rm CI}(\varphi)]$ of eqs.~\eqn{DhatphiFO} and \eqn{DhatphiCI} for
a summation of the perturbative series up to four different orders, namely the
4th (dotted line), 5th (dashed-double-dotted line), 6th (dashed-dotted line)
and 7th (dashed line) order. The input for the QCD coupling was chosen to be
$\as(M_\tau)=0.3156$. Also displayed as the solid line is the Borel sum
$\RE[\wh D_{\rm BS}(\varphi)]$ according to our physical model presented in
section~\ref{sect6}.\label{fig9}}}

In figure~\ref{fig9}, we display a graphical account of our numerical results
for the real parts $\RE[\wh D_{\rm FO}(\varphi)]$ and
$\RE[\wh D_{\rm CI}(\varphi)]$ in the range $0<\varphi<2\pi$ and for
$\as(M_\tau)=0.3156$. We have drawn four different lines corresponding to
a truncation of the perturbative series at the 4th (dotted line), 5th
(dashed-double-dotted line), 6th (dashed-dotted line) and 7th (dashed line)
order. In addition, the solid line corresponds to a resummation of the
perturbative series according to the Borel sum of the physical model for the
Adler function introduced in section~\ref{sect6}. The required perturbative
coefficients $c_{6,1}$ and $c_{7,1}$ for this model are given in
table~\ref{tab2}. The reason for showing the summation of the series up to
$n=7$ lies in the fact that the smallest summand of both series is found in
this range. 

Various observations can be made on the basis of figure~\ref{fig9}. First of
all, as can also be seen from eqs.~\eqn{DhatphiFO} and \eqn{DhatphiCI}, at the
euclidian point $\varphi=\pi$ on the negative real $s$-axis, $\wh D_{\rm FO}$
and $\wh D_{\rm CI}$ are identical. The closest approach of the partial sums
to the Borel sum for $\varphi=\pi$ is reached at the 6th order (compare
figure~\ref{figadler}). Close to the minkowskian region (the positive real
$s$-axis), that is $\varphi$ near zero or $2\pi$, FOPT converges rather slowly,
and only at the 7th order, the series approaches the shape displayed by the
Borel sum. The slow convergence of FOPT close to the minkowskian region has
already been noted in ref.~\cite{ddhmz08}, and was taken as an argument that
CIPT should be preferable. Indeed, as is obvious from the lower plot of
figure~\ref{fig9}, CIPT converges very nicely for all angles and mostly even
better for $\varphi$ not near $\pi$ than at the euclidian point. However,
the resulting sums do not at all approach the shape of the Borel sum. Thus,
except at a few isolated points, CIPT for the Adler function in the complex
plane exhibits a similar problem as for $R_\tau$, shown in figure~\ref{fig6},
that the series exhibits apparent convergence and stability, but the resultant
value is much further from the ``true result'' than any theoretical error 
estimate would suggest. For the Adler function CIPT lies below the Borel sum in
the central region and above elsewhere, while for $R_\tau$  the weight function
$(1-x)^3(1+x)$ with $x={\rm e}^{i\varphi}$ in the integration over the angle
$\varphi$ implies that $\delta^{(0)}_{\rm CI}$ is below the Borel sum
$\delta^{(0)}_{\rm BS}$ as found in section~\ref{sect6}. On the other hand,
FOPT turns out to be much closer to the Borel sum, since the problematic region
close to the minkowskian axis is strongly suppressed in $R_\tau$ by the triple
zero of the kinematical weight function at $s=M_\tau^2$. The problems with CIPT
for the Adler function in the complex plane arise essentially for the same
reason as for $R_\tau$. In FOPT at each order we observe sizeable cancellations
between the independent coefficients $c_{n,1}$ and the running effects which
are missed by CIPT, although here, due to the angular dependence, the situation
is less transparent.

\bigskip

\providecommand{\href}[2]{#2}\begingroup\raggedright\endgroup

\end{document}